\documentclass[journal]{IEEEtran}
\usepackage[dvips]{graphicx}
\usepackage{epsfig}
\usepackage{amsmath}
\usepackage{epsfig}
\usepackage{array}
\usepackage{amssymb}
\usepackage{color}

\newtheorem{Theorem}{\textbf{Theorem}}
\newtheorem{Corollary}{\textbf{Corollary}}

\begin{document}

\title{Secret-Message Transmission by Echoing Encrypted Probes --- STEEP}

\author{Yingbo Hua\thanks{Department of Electrical and Computer Engineering,
University of California, Riverside, CA 92521, USA. Email: yhua@ece.ucr.edu. This work was supported in part by the Department of Defense under W911NF-20-2-0267. The views and conclusions contained in this
document are those of the author and should not be interpreted as representing the official policies, either
expressed or implied, of  the U.S. Government. The U.S. Government is
authorized to reproduce and distribute reprints for Government purposes notwithstanding any copyright
notation herein.}
}

\maketitle

\begin{abstract}
This paper examines the properties of the lower and upper bounds established by Maurer, Ahlswede and Csiszar (MAC) for secret-key capacity in the case of channel probing over single-input and single-output (SISO) channels. Inspired by the insights into MAC's bounds, a scheme called secret-message transmission by echoing encrypted probes (STEEP) is proposed. STEEP consists of two phases: in phase 1, Alice sends random probes over a probing channel to Bob; in phase 2, Bob echoes back an estimated version of the probes, but encrypted by a secret, over a high-quality return channel. Provided that Eve is unable to obtain the exact probes transmitted by Alice in phase 1, STEEP guarantees a positive secrecy rate from Bob to Alice over the return channel even if Eve's channel strength during channel probing is stronger than Bob's. STEEP is applicable to both physical layer and upper layers in connected networks.
\end{abstract}

\section{Introduction}

The design of how to transmit a secret message through a network of nodes (including people, agents, devices, machines and organizations) has attracted human's interest for ages, and its development from the information-theoretic (IT) perspectives was pioneered by Shannon \cite{Shannon1948} in 1940s. The IT achievable limits based on one-way transmission over a wire-tap channel (WTC) model was established in 1970s by Wyner \cite{Wyner1975}, and Csiszar and Korner \cite{Csiszar1978}. Further developments of coding techniques and their theoretical bounds for one-way transmission are well documented in \cite{Bloch2008}, \cite{Yang2019} and many references therein.  The achievable secrecy capacity (in bits per channel use in each coherence period) of one-way transmission is known to be $\xi_s=\xi_m-\xi_e$ which is zero if the main channel capacity $\xi_m$ (from Alice to Bob) is less than or equal to the eavesdropping channel capacity $\xi_e$ (from Alice to Eve). For  finite-length packets, a smallest penalty to $\xi_s$ was recently achieved in $\cite{Yang2019}$, and many
 applications of such achievable secrecy capacity can be found via \cite{Ari2022} and \cite{Feng2022}.

For a vast variety of situations in the real world, the transmission of a secret does not need to be constrained to be one-way. Cooperative two-way transmissions between two agents in a modern network are widely feasible. Indeed, in parallel to the WTC based developments, there is another branch of developments called secret key generation (SKG) via public communications, which was pioneered by Maurer in  \cite{Maurer1993} and Ahlswede and Csiszar in \cite{Ahlswede1993} in 1990s. One of the widely applicable results developed by Maurer, Ahlswede and Csiszar (MAC) is thoroughly revisited by Bloch and Barros in \cite{Bloch2008}, which can be stated as follows:

Let the data sets available to Alice, Bob and Eve be respectively $\mathcal{A}$, $\mathcal{B}$ and $\mathcal{E}$. Then there is a public communication scheme such that the achievable secret-key capacity $C_{key}$ (in bits per independent realization of $\mathcal{A}$, $\mathcal{B}$ and $\mathcal{E}$) between Alice and Bob against Eve is bounded by
\begin{equation}\label{eq:MAC}
  C_L\leq C_{key}\leq C_U
\end{equation}
with $C_L=\max(C_A,C_B)$, $C_A=I(\mathcal{A};\mathcal{B})-I(\mathcal{A};\mathcal{E})$, $C_B=I(\mathcal{A};\mathcal{B})-I(\mathcal{B};\mathcal{E})$,  $C_U=\min(I(\mathcal{A};\mathcal{B}),C_E)$, and $C_E=I(\mathcal{A};\mathcal{B}|\mathcal{E})$. Here $I(A;B|C)$ denotes the mutual information between $A$ and $B$ conditioned on $C$. We will refer to $C_L$ and $C_U$ as MAC's lower and upper bounds respectively.

However, despite the fact that any secret key of $n$ bits established between Alice and Bob allows one of them to transmit a message of $n$ bits to the other in complete secrecy, the past works on SKG and/or MAC's bounds have been largely treated in isolation from WTC based secret-message transmission. Such examples include \cite{Wilson2007}, \cite{Wallace2010}, \cite{HessamMahdavifar2020}, \cite{Li2022} as well as our own recent works \cite{Hua2023} and \cite{HuaMaksud2023March}.

In this work, we will look deeper into some special cases of a channel probing scheme studied in \cite{Hua2023}. Inspired by the insights into MAC's bounds of the scheme, we will propose a scheme called ``secret-message transmission by echoing encrypted probes (STEEP)''. Specifically, STEEP consists of two phases:

Phase 1: Alice transmits probing signals (or probes) to Bob, and most likely and unintentionally to Eve, over  one or more probing channels, from which Eve must obtain a noisy version of the probes while Bob may receive a noisier version of the probes.

Phase 2: Bob echoes back his estimates of the probes, but encrypted by his secret, over one or more return channels that have much higher quality than the probing channels in phase 1. This creates an effective WTC model from Bob to Alice and Eve in such a way that the effective return channel from Bob to Alice is guaranteed to be stronger than that from Bob to Eve. Consequently, any established WTC scheme can be applied here to achieve a positive secrecy rate from Bob to Alice.

STEEP allows Alice to receive a secret message from Bob reliably, and it yields a positive secrecy rate as long as Eve observes a noisy (as opposed to noiseless) version of the probes. STEEP provides an important unification of the prior theories and algorithms developed for WTC and SKG. There is little restriction on the probing channel and/or the return channel. For example, the probing channel can be wireless or wireline, single-hop or multiple-hop, analog or digital, fading or non-fading, etc. And the return channel can be viewed as any public channel. Unlike the scheme in \cite{Bloch2008} where a one-way scheme is developed, STEEP is a two-way scheme. Unlike the scheme in \cite{Maurer1993} where many iterative transmissions between Alice and Bob via public channel are assumed, STEEP only needs one round-trip communication. Unlike \cite{Feng2022} where the achieved secrecy rate is zero if Eve's channel from Alice is stronger than Bob's channel from Alice, STEEP yields a positive secrecy rate even if Eve's channel from Alice (during channel probing) is stronger than that of Bob's channel from Alice.
 Unlike \cite{HessamMahdavifar2020} and \cite{Li2022} where public pilots are avoided in hope to achieve a positive secrecy that grows with the number of probing symbols in each coherence period, STEEP is guaranteed to yield a secrecy that increases with the number of probing symbols even for a static probing channel and even with public pilots used for channel estimation. Unlike \cite{Lai2012} which requires a perfectly-reciprocal fading channel between users and a weak channel at Eve for a SKG framework, STEEP can yield a positive secrecy rate even if users' channel gain is constant, non-reciprocal, known to Eve, and/or weaker than Eve's channel gain.
Unlike all prior works constrained to the physical layer, STEEP applies to all layers of networks.

Ignoring the contribution from correlated channel gains between Alice and Bob, the achievable secrecy limit (in bits per probing sample) of STEEP for analog probing channels is
\begin{equation}\label{eq:xi}
  \xi_{\texttt{STEEP},\texttt{AC}}=\mathbb{E}\left \{\log\left (1+\frac{\texttt{SNR}_{B,A}}{1+\texttt{SNR}_{E,A}}\right )\right\}
\end{equation}
where $\texttt{SNR}_{B,A}$ is Bob's instantaneous signal-to-noise ratio (SNR) during channel probing, and $\texttt{SNR}_{E,A}$ is Eve's (after matched filtering in the case of multiple antennas on Eve). The expectation $\mathbb{E}$ can be dropped for a probing channel with long coherence time. We see that STEEP can deliver a positive secrecy rate virtually for any $\texttt{SNR}_{E,A}$. This should make STEEP attractive in many applications where one might only know an upper bound of $\texttt{SNR}_{E,A}$, which could be larger than $\texttt{SNR}_{B,A}$.

The development of STEEP is based on important insights into MAC's bounds of channel probing for SKG in the context of a single-input and single-out (SISO) channel between Alice and Bob. We will first present these insights in sections \ref{sec:two-way}, \ref{sec:return}, \ref{sec:pruned} and \ref{sec:estimation_of_s}. Specifically, section \ref{sec:two-way} presents MAC's lower and upper bounds based on a (half-duplex) two-way channel probing scheme, which then leads to an important observation that for one-way channel probing, MAC's lower and upper bounds meet each other and the  corresponding secret-key capacity $C_{key}$ is a positive and linearly increasing function of the number of probing symbols even if Eve's channel from Alice (during probing) is stronger than that of Bob's channel from Alice. In section \ref{sec:return}, we show that after channel probing from Alice to Bob, if Bob sends a combination of a random sequence (of his own) and his received (noisy or noiseless) probes back to Alice (and also to Eve) via a high-quality return channel, MAC's lower bound converges to $C_{key}$ when a large power is used by Bob for the random sequence. In section \ref{sec:pruned}, we present a refined insight into MAC's lower bound where Bob discards his received probes after he has generated and transmitted the random sequence in combination with his received probes. In section \ref{sec:estimation_of_s}, we focus on the optimal estimation by Alice and Eve of the random sequence transmitted by Bob, from which an important advantage of the round-trip communication (i.e., channel probing in one direction followed by transmission of encrypted probes in the other direction) becomes more clear.

With the above insights, in section \ref{sec:STEEP_analog}, we present our scheme called STEEP for analog channels.  In section \ref{sec:STEEP_digital}, we show that the principle of STEEP is also applicable to digital channels in any connected networks. The only fundamental requirement in order for STEEP to yield a positive secrecy rate is that Eve does not receive the exact probing symbols sent by Alice.

\section{Two-Way SISO Channel Probing}\label{sec:two-way}

Two-way (half-duplex) channel probing schemes for SKG through a MIMO channel between Alice and Bob were recently studied in \cite{Hua2023} where the degree of freedom of MAC's bounds relative to the probing power is shown. But the complexity caused by the MIMO channel still poses a challenge to fully understand the MIMO based MAC's bounds. Some latest development on the MIMO case is shown in \cite{MaksudHua2023Nov}.

In this paper, we will focus on the SISO channel between Alice and Bob in order to understand the exact MAC's bounds and their implications. But we assume that Eve has $n_E\geq 1$ antennas. For a wireline network, the multiple antennas on Eve would correspond to multiple tapping points on the network. We will only consider (discrete-time) analog channels (i.e., physical layer channels) until section \ref{sec:STEEP_digital}.

Furthermore, we assume that Alice has applied a public pilot (of sufficient power) so that Bob has obtained his receive channel response $h_{B,A}$, and Eve has obtained her channel response vector $\mathbf{g}_A\in\mathcal{C}^{n_E\times 1}$ (relative to Alice). Similarly, Bob has applied a public pilot so that Alice knows her receive channel response $h_{A,B}$, and Eve knows her channel response vector $\mathbf{g}_B\in\mathcal{C}^{n_E\times 1}$ (relative to Bob).

Then Alice sends $m_A$ i.i.d. random probing symbols (only known to Alice), denoted by $X_A=\{x_A(k),k=1,\cdots,m_A\}$, over the SISO probing channel to Bob. After that, Bob also sends $m_B$ i.i.d. random probing symbols $X_B=\{x_B(k),k=1,\cdots,m_B\}$ (only known to Bob) to Alice over the SISO probing channel (in reverse direction). Correspondingly, Alice and Bob receive respectively
\begin{align}\label{}
  &y_A(k)=h_{A,B} x_B(k)+ w_B(k),\\
  &y_B(k)=h_{B,A} x_A(k)+ w_A(k).
\end{align}
And Eve receives both of the following:
\begin{equation}\label{eq:eAk}
  \mathbf{e}_A(k)= \mathbf{g}_A x_A(k) + \mathbf{w}_{E,A}(k),
\end{equation}
\begin{equation}\label{eq:eBk}
  \mathbf{e}_B(k)= \mathbf{g}_B x_B(k) + \mathbf{w}_{E,B}(k).
\end{equation}
Here all complex components of $x_A(k)$, $x_B(k)$, $w_A(k)$, $w_B(k)$, $\mathbf{w}_{E,A}(k)$ and $\mathbf{w}_{E,B}$ are i.i.d. circular complex Gaussian with zero mean and their variances denoted by
$p_A$, $p_B$, $\sigma_A^2$, $\sigma_B^2$, $\sigma_{E,A}^2$ and $\sigma_{E,B}^2$ respectively.

In this section, we also assume that
$h_{A,B}$ is only known to Alice;
$h_{B,A}$ is only known to Bob;
  $\mathbf{g}_A$ and $\mathbf{g}_B$ are only known to Eve;
$h_{A,B}$ and $h_{B,A}$ are jointly Gaussian with zero mean and the covariance matrix
  $\left [\begin{array}{cc}
  1 & \rho \\
 \rho^* & 1
 \end{array}
 \right ]$. Here $\rho=\mathbb{E}\{h_{A,B}h_{B,A}^*\}$ and $|\rho|<1$. Also in this section, $\{h_{A,B},h_{B,A}\}$ is independent of $\{\mathbf{g}_A,\mathbf{g}_B\}$.

To apply MAC's bounds, we let $\mathcal{A}=\{\mathcal{A}_1,h_{A,B}\}$ with $\mathcal{A}_1=\{X_A,Y_A\}$ and $Y_A=\{y_A(k),k=1,\cdots,m_B\}$;
$\mathcal{B}=\{\mathcal{B}_1,h_{B,A}\}$ with $\mathcal{B}_1=\{X_B,Y_B\}$ and $Y_B=\{y_B(k),k=1,\cdots,m_A\}$;
 $\mathcal{E}=\{\mathcal{E}_1,\mathbf{g}_A,\mathbf{g}_B\}$ with $\mathcal{E}_1=\{E_A,E_B\}$, $E_A=\{\mathbf{e}_A(k),k=1,\cdots,m_A\}$ and $E_B=\{\mathbf{e}_B(k),k=1,\cdots,m_B\}$.

\begin{Theorem}\label{Theorem1}
Based on the above model of $\{\mathcal{A}, \mathcal{B}, \mathcal{E}\}$, MAC's bounds shown in \eqref{eq:MAC} are governed by
   \begin{align}\label{eq:CA3}
 &C_A=\alpha+m_B\xi_{A,B}+m_A\gamma_{B,A},\\
\label{eq:CB3}
 &C_B=\alpha+m_A\xi_{B,A}+m_B\gamma_{A,B},\\
\label{eq:CE3}
 &C_E=\alpha+m_A\xi_{B,A}+m_B\xi_{A,B},
 \end{align}
 where $\alpha=-\log(1-|\rho|^2)$,
 \begin{equation}\label{}
   \xi_{B,A}=\mathbb{E}\left \{\log\left (1+\frac{p_A|h_{B,A}|^2/\sigma_B^2}{p_A\|\mathbf{g}_A\|^2/\sigma_{E,A}^2+1}\right )\right \},
 \end{equation}
 \begin{equation}\label{}
   \gamma_{B,A}=\mathbb{E}\left \{\log\frac{p_A|h_{B,A}|^2/\sigma_B^2+1}
 {p_A\|\mathbf{g}_A\|^2/\sigma_{E,A}^2+1}\right \},
 \end{equation}
 and $\xi_{A,B}$ and $\gamma_{A,B}$ are defined accordingly by exchanging ``$A$'' and ``$B$''.
\end{Theorem}

The proof is provided after the discussion shown next.

 \subsection{Discussion of Theorem \ref{Theorem1}}\label{sec:discussion_theorem1}

 We will also use
 \begin{equation}\label{eq:phiA}
  \phi_{A,B}=\frac{p_B|h_{A,B}|^2/\sigma_A^2}{p_B\|\mathbf{g}_B\|^2/\sigma_{E,B}^2+1},
\end{equation}
\begin{equation}\label{eq:phiB}
  \phi_{B,A}=\frac{p_A|h_{B,A}|^2/\sigma_B^2}{p_A\|\mathbf{g}_A\|^2/\sigma_{E,A}^2+1},
\end{equation}
so that $\xi_{A,B}=\mathbb{E}\{\log(1+\phi_{A,B})\}$ and $\xi_{B,A}=\mathbb{E}\{\log(1+\phi_{B,A})\}$.

 It is obvious that $\xi_{A,B}>0$, $\xi_{B,A}>0$, $\gamma_{A,B}<\xi_{A,B}$ and $\gamma_{B,A}<\xi_{B,A}$. Also $\gamma_{B,A}>0$ if and only if the probing channel from Alice to Bob is stronger than that from Alice to Eve, i.e., $|h_{B,A}|^2/\sigma_B^2>\|\mathbf{g}_A\|^2/\sigma_{E,A}^2$. Similarly, $\gamma_{A,B}>0$ if and only if the probing channel from Bob to Alice is stronger than that from Bob to Eve, i.e., $|h_{A,B}|^2/\sigma_A^2>\|\mathbf{g}_B\|^2/\sigma_{E,B}^2$. In the absence of the knowledge of whether Eve's channel is weak or not, we can guarantee $C_A>0$ by choose $m_A=0$. Similarly, we can guarantee $C_B>0$ by choose $m_B=0$. Furthermore, with $m_A=0$, we have $C_A=C_E$ which is then $C_{key}$. And with $m_B=0$, we have $C_B=C_E$ which is then $C_{key}$. Note that it follows from \eqref{eq:MAC} that $\max(C_A,C_B)\leq C_{key}\leq C_E$.

 \begin{Corollary}\label{Corollary1}
   MAC's lower and upper bounds for one-way channel probing coincide with each other. Specifically, if $m_A>m_B=0$,
   \begin{equation}\label{eq:CB_one_way}
     C_{key}=C_B=C_E=\alpha+m_A\xi_{B,A}
   \end{equation}
    which is positive and increases linearly with $m_A$. Similarly, if $m_B>m_A=0$,
    \begin{equation}\label{eq:CA_one_way}
      C_{key}=C_A=C_E=\alpha+m_B\xi_{A,B}
    \end{equation}
     which is positive and increases linearly with $m_B$.
 \end{Corollary}

 Here $m_A\xi_{B,A}$ in \eqref{eq:CB_one_way} is the amount of secrecy achievable from $m_A$ probing samples from Alice. So, we can refer to $\xi_{B,A}$ as secrecy rate in bits per probing sample from Alice. Similarly, $\xi_{A,B}$ can be referred to as secrecy rate in bits per probing sample from Bob.

  It should be of interest to note that if the two-way channel probing scheme can be (and is) conducted in full-duplex (i.e., at the same frequency and the same time) subject to the same channel model at Eve (i.e., \eqref{eq:eAk} and \eqref{eq:eBk}), Corollary \ref{Corollary1} with $n_E=1$ would coincide exactly with the lower bound of achievable secrecy rate shown in Proposition 1 in \cite{Khisti2012} (subject to the pilot power being much larger than the power of the probing symbols). More specifically, with or without full-duplex, the first term $\alpha$ in \eqref{eq:CB_one_way} and \eqref{eq:CA_one_way} represents the total secrecy contributed by the correlation between $h_{A,B}$ and $h_{B,A}$. In the case of full-duplex, $\alpha$ is achieved using a single sampling interval for both public pilots from Alice and Bob.
   Also, for full-duplex two-way probing, the second terms in \eqref{eq:CB_one_way} and \eqref{eq:CA_one_way} (under ``$m_B=m$ and $m_A=0$'' and ``$m_A=m$ and $m_B=0$'' respectively) should be added together as an achievable secrecy due to the $m$ sampling intervals for the random symbols from Alice and Bob. This is because under full-duplex two-way probing, the data sets $\mathcal{S}=\{\mathcal{A},\mathcal{B},\mathcal{E}\}$ collected by Alice, Bob and Eve can be partitioned as $\mathcal{S}=\{\mathcal{S}_A,\mathcal{S}_B\} $ with $\mathcal{S}_A=\{\mathcal{A}_A,\mathcal{B}_A,\mathcal{E}_A\}$ and $\mathcal{S}_B=\{\mathcal{A}_B,\mathcal{B}_B,\mathcal{E}_B\}$. Here (not counting the parts associated with public pilots), $\mathcal{S}_A$ is associated with the transmission from Alice, and $\mathcal{S}_B$ is associated with the transmission from Bob. Hence $\mathcal{S}_A$ is independent of $\mathcal{S}_B$ when conditioned on channel state information. More interestingly, this same reasoning along with Corollary \ref{Corollary1} implies that the achievable lower bound shown in Proposition 1 in \cite{Khisti2012} is also the achievable upper bound.

 However, unlike \cite{Khisti2012} which focuses on SKG from a full-duplex SISO channel, this paper considers a half-duplex SISO (probing) channel, and our proof of secret-key capacity is based on the established MAC's bounds. We will not need to re-establish the correctness and/or tightness (or any asymptotical properties) of MAC's  bounds.  MAC's bounds are somewhat universal and applicable to the data sets $\mathcal{A}$, $\mathcal{B}$ and $\mathcal{C}$ formed under each of our considered cases. Furthermore, this paper is not only about the achievable secrecy rate but also about a simple scheme called STEEP to be shown.

 Because of Corollary \ref{Corollary1} and the need to guarantee a positive secrecy rate under any conditions of Eve's channel, we will next focus on one-way channel probing for secret-message transmission. A goal in this paper is to present a simple method (simpler than the back-and-forth public communication scheme shown in \cite{Maurer1993}) to transmit a secret between Alice and Bob with the secrecy capacity approaching that shown in Corollary \ref{Corollary1}.

 \subsection{Proof of Theorem \ref{Theorem1}}

Let us start with $C_A=I(\mathcal{A};\mathcal{B})-I(\mathcal{A};\mathcal{E})$ and analyze $I(\mathcal{A};\mathcal{B})$ and $I(\mathcal{A};\mathcal{E})$ separately as follows.

\subsubsection{Analysis of $I(\mathcal{A};\mathcal{B})$}
We know
\begin{align}\label{eq:IAB}
&I(\mathcal{A};\mathcal{B})=I(h_{A,B};\mathcal{B})+I(\mathcal{A}_1;\mathcal{B}|h_{A,B})\notag\\
&=I(h_{A,B};h_{B,A})+I(h_{A,B};\mathcal{B}_1|h_{B,A})\notag\\
&\,\,+I(\mathcal{A}_1;h_{B,A}|h_{A,B})+
I(\mathcal{A}_1;\mathcal{B}_1|h_{B,A},h_{A,B}).
\end{align}
Here,
 \begin{align}\label{eq:IhABhBA}
 &I(h_{A,B};h_{B,A}) = h(h_{A,B})-h(h_{A,B}|h_{B,A}) \notag\\
 &= \log(e\pi)-\log(e\pi(1-|\rho|^2))
 =-\log(1-|\rho|^2)
 \end{align}
 where $1-|\rho|^2$ is the variance of $h_{A,B}$ when conditioned on $h_{B,A}$.
 And
 \begin{align}
 &I(h_{A,B};\mathcal{B}_1|h_{B,A}) = I(h_{A,B};Y_B|X_B,h_{B,A})\notag\\
 &=h(Y_B|X_B,h_{B,A})-h(Y_B|h_{A,B},X_B,h_{B,A})\notag\\
 &=h(Y_B|h_{B,A})-h(Y_B|h_{B,A})=0.
 \end{align}
 By symmetry, we also have $I(\mathcal{A}_1;h_{B,A}|h_{A,B})=0$.

 For the 4th term in \eqref{eq:IAB},
 we will write $I(\mathcal{A}_1;\mathcal{B}_1|h_{B,A},h_{A,B})=I(\mathcal{A}_1;\mathcal{B}_1|h_T)$. It follows that
 \begin{align}
 &I(\mathcal{A}_1;\mathcal{B}_1|h_T)=I(X_A;\mathcal{B}_1|h_T)
 +I(Y_A;\mathcal{B}_1|X_A,h_T)\notag\\
 &=I(X_A;Y_B|h_T)+I(Y_A;X_B|h_T),
 \end{align}
 where we have used the independence between $X_A$ and $X_B$. Furthermore,
 \begin{align}\label{eq:IA1B1hT}
 &I(\mathcal{A}_1;\mathcal{B}_1|h_T)=h(Y_B|h_T)-h(Y_B|X_A,h_T)\notag\\
 &\,\, +h(Y_A|h_T)-h(Y_A|X_B,h_T)\notag\\
 &=m_A\mathbb{E}\{\log(p_A|h_{B,A}|^2/\sigma_B^2+1)\}\notag\\
 &\,\,
 +m_B\mathbb{E}\{\log(p_B|h_{A,B}|^2/\sigma_A^2+1)\}.
 \end{align}

 Combining the above results in \eqref{eq:IAB} yields
 \begin{align}\label{eq:IAB2}
&I(\mathcal{A};\mathcal{B})=-\log(1-|\rho|^2)+m_A\mathbb{E}\{\log(p_A|h_{B,A}|^2/\sigma_B^2+1)\}
\notag\\
&\,\, +m_B\mathbb{E}\{\log(p_B|h_{A,B}|^2/\sigma_A^2+1)\}.
\end{align}

\subsubsection{Analysis of $I(\mathcal{A};\mathcal{E})$}
It follows that
 \begin{align}
 &I(\mathcal{A};\mathcal{E})=I(\mathcal{A}_1;\mathcal{E}|h_{A,B})
 =I(\mathcal{A}_1;\mathcal{E}_1|h_{A,B},\mathbf{g}_A,\mathbf{g}_B)
 \end{align}
 where we have applied the independent between $h_{A,B}$ and $\{\mathbf{g}_A,\mathbf{g}_B\}$. We will now write $I(\mathcal{A}_1;\mathcal{E}_1|h_{A,B},\mathbf{g}_A,\mathbf{g}_B)=
 I(\mathcal{A}_1;\mathcal{E}_1|c_A)$. It follows that
 \begin{align}
&I(\mathcal{A};\mathcal{E})=I(\mathcal{A}_1;\mathcal{E}_1|c_A)\notag\\
&=I(X_A;\mathcal{E}_1|c_A)+I(Y_A;\mathcal{E}_1|X_A,c_A)\notag\\
&=I(X_A;E_A|c_A)+I(Y_A;E_B|X_A,c_A),
 \end{align}
 where we have used the independence between $X_A$ and $E_B$, and the independence between $Y_A$ and $E_A$. Furthermore, $I(Y_A;E_B|X_A,c_A)=I(Y_A;E_B|c_A)$ due to independence between $X_A$ and $\{Y_A,E_B\}$. It follows that
  \begin{align}\label{eq:IAE}
&I(\mathcal{A};\mathcal{E})=I(X_A;E_A|c_A)+I(Y_A;E_B|c_A).
 \end{align}
 Here
 \begin{align}\label{eq:IXAEAcA}
   &I(X_A;E_A|c_A)=h(E_A|c_A)-h(E_A|X_A,c_A)\notag\\
   &=
   m_A\mathbb{E}\{\log|p_A\mathbf{g}_A\mathbf{g}_A^H/\sigma_{E,A}^2
   +\mathbf{I}_{n_E}|\}\notag\\
   &=m_A\mathbb{E}\{\log (p_A\|\mathbf{g}_A\|^2/\sigma_{E,A}^2
   +1)\},
 \end{align}
 and, with $\mathbf{D}_A=diag(\sigma_A^2,\sigma_{E,B}^2,\cdots,\sigma_{E,B}^2)
 =diag(\sigma_A^2,\sigma_{E,B}^2\mathbf{I}_{n_E})$  and $\mathbf{g}_B'=[h_{A,B},\mathbf{g}_B^T]^T$,
 \begin{align}\label{eq:IXAEBcA}
&I(Y_A;E_B|c_A)=h(E_B|c_A)-h(E_B|Y_A,c_A)\notag\\
&=h(E_B|c_A)+h(Y_A|c_A)-h(E_B,Y_A|c_A)\notag\\
&=m_B\mathbb{E}\{\log|p_B\mathbf{g}_B\mathbf{g}_B^H+\sigma_{E,B}^2\mathbf{I}_{n_E}|\}\notag\\
&\,\,+m_B\mathbb{E}\{\log(p_B|h_{A,B}|^2+\sigma_A^2)\notag\\
&\,\,-m_B\mathbb{E}\{\log|p_B\mathbf{g}_B'\mathbf{g}_B'^H+\mathbf{D}_A|\}\notag\\
&=m_B\mathbb{E}\{\log(\sigma_{E,B}^{2n_E}(p_B\|\mathbf{g}_B\|^2/\sigma_{E,B}^2+1))\}\notag\\
&\,\,+m_B\mathbb{E}\{\log(\sigma_A^2(p_B|h_{A,B}|^2/\sigma_A^2+1))\notag\\
&\,\,-m_B\mathbb{E}\{\log(\sigma_A^2\sigma_{E,B}^{2n_E}(p_B|h_{A,B}|^2/\sigma_A^2\notag\\
&\,\,+
p_B\|\mathbf{g}_B\|^2/\sigma_{E,B}^2
+1))\}\notag\\
&=m_B\mathbb{E}\{\log(p_B\|\mathbf{g}_B\|^2/\sigma_{E,B}^2+1)\}\notag\\
&\,\,+m_B\mathbb{E}\{\log(p_B|h_{A,B}|^2/\sigma_A^2+1)\notag\\
&\,\,-m_B\mathbb{E}\{\log(p_B|h_{A,B}|^2/\sigma_A^2+
p_B\|\mathbf{g}_B\|^2/\sigma_{E,B}^2
+1)\}.
 \end{align}
 Note that we have used $|p_B\mathbf{g}_B'\mathbf{g}_B'^H+\mathbf{D}_A|=|\mathbf{D}_A|\cdot |p_B\mathbf{D}_A^{-1/2}\mathbf{g}_B'\mathbf{g}_B'^H\mathbf{D}_A^{-1/2}+\mathbf{I}_{n_E+1}|
 =|\mathbf{D}_A|(p_B\mathbf{g}_B'^H\mathbf{D}_A^{-1}\mathbf{g}_B'+1)$.

 Combining the above results in \eqref{eq:IAE} yields
 \begin{align}\label{eq:IAE2}
&I(\mathcal{A};\mathcal{E})=m_A\mathbb{E}\{\log (p_A\|\mathbf{g}_A\|^2/\sigma_{E,A}^2
   +1)\}\notag\\
   &\,\,+m_B\mathbb{E}\{\log(p_B\|\mathbf{g}_B\|^2/\sigma_{E,B}^2+1)\}\notag\\
&\,\,+m_B\mathbb{E}\{\log(p_B|h_{A,B}|^2/\sigma_A^2+1)\notag\\
&\,\,-m_B\mathbb{E}\{\log(p_B|h_{A,B}|^2/\sigma_A^2+
p_B\|\mathbf{g}_B\|^2/\sigma_{E,B}^2
+1)\}.
 \end{align}

Combining \eqref{eq:IAB2} and \eqref{eq:IAE2} yields \eqref{eq:CA3}.
By symmetry between ``$A$'' and ``$B$'', \eqref{eq:CB3} follows from \eqref{eq:CA3}.

 \subsubsection{Analysis of $C_E$}

 We know
 \begin{align}\label{eq:CE}
 &C_E=I(\mathcal{A};\mathcal{B}|\mathcal{E})\notag\\
 &=I(h_{A,B};\mathcal{B}|\mathcal{E})+I(\mathcal{A}_1;\mathcal{B}|h_{A,B},\mathcal{E})\notag\\
 &=I(h_{A,B};h_{B,A})+I(h_{A,B};\mathcal{B}_1|h_{B,A},\mathcal{E})\notag\\
 &\,\,
 +I(\mathcal{A}_1;h_{B,A}|h_{A,B},\mathcal{E})+
 I(\mathcal{A}_1;\mathcal{B}_1|h_{B,A},h_{A,B},\mathcal{E}).
 \end{align}
Here $I(h_{A,B};h_{B,A})$ is given by \eqref{eq:IhABhBA}. The 2nd term in \eqref{eq:CE} is $I(h_{A,B};\mathcal{B}_1|h_{B,A},\mathcal{E})=I(h_{A,B};Y_B|X_B,h_{B,A},\mathcal{E})=
h(Y_B|X_B,h_{B,A},\mathcal{E})-h(Y_B|h_{A,B},X_B,h_{B,A},\mathcal{E})
=h(Y_B|X_B,h_{B,A},E_A,E_B)-h(Y_B|h_{A,B},X_B,h_{B,A},E_A,E_B)
=h(Y_B|X_B,h_{B,A},E_A,E_B)-h(Y_B|X_B,h_{B,A},E_A,E_B)=0$. Similarly, the 3rd term in \eqref{eq:CE} is also zero.

We will use $h_T$ to denote $\{h_{A,B},h_{B,A}\}$, and $c_T$ to denote $\{h_{A,B},h_{B,A},\mathbf{g}_A,\mathbf{g}_B\}$. Then the 4th term in \eqref{eq:CE} is $I(\mathcal{A}_1;\mathcal{B}_1|h_T,\mathcal{E})=I(\mathcal{A}_1;\mathcal{B}_1|h_T,
 \mathbf{g}_A,\mathbf{g}_B,\mathcal{E}_1)=I(\mathcal{A}_1;\mathcal{B}_1|c_T,\mathcal{E}_1)$. It follows that
 \begin{align}\label{eq:IA1B1cTE}
 &I(\mathcal{A}_1;\mathcal{B}_1|c_T,\mathcal{E})=I(X_A;\mathcal{B}_1|c_T,\mathcal{E}_1)
 +I(Y_A;\mathcal{B}_1|X_A,c_T,\mathcal{E}_1)\notag\\
 &=I(X_A;X_B|c_T,\mathcal{E}_1)+I(X_A;Y_B|X_B,c_T,\mathcal{E}_1)\notag\\
 &\,\,
 +I(Y_A;X_B|X_A,c_T,\mathcal{E}_1)+I(Y_A;Y_B|X_B,X_A,h_T,\mathcal{E}_1)\notag\\
 &=T_1+T_2+T_3+T_4.
 \end{align}

 The first term in \eqref{eq:IA1B1cTE} is
 \begin{align}
 &T_1=h(X_A|c_T,\mathcal{E}_1)-h(X_A|X_B,c_T,\mathcal{E}_1)\notag\\
 &=h(\mathcal{E}_1|X_A,c_T)+h(X_A|c_T)-h(\mathcal{E}_1|c_T)\notag\\
 &\,\,-[h(\mathcal{E}_1|X_A,X_B,c_T)+h(X_A|X_B,c_T)-h(\mathcal{E}_1|X_B,c_T)]\notag\\
 &=h(\mathcal{E}_1|X_A,c_T)-h(\mathcal{E}_1|c_T)-h(\mathcal{E}_1|X_A,X_B,c_T)\notag\\
 &\,\,+
 h(\mathcal{E}_1|X_B,c_T)],
 \end{align}
 where $h(\mathcal{E}_1|X_A,c_T)=h(E_A|X_A,c_T)+h(E_B|E_A,X_A,c_T)=
 h(E_A|X_A,c_T)+h(E_B|c_T)$, $h(\mathcal{E}_1|c_T)=h(E_A|c_T)+h(E_B|E_A,c_T)
 =h(E_A|c_T)+h(E_B|c_T)$, and $h(\mathcal{E}_1|X_B,c_T)=h(E_A|X_B,c_T)+h(E_B|E_A,X_B,c_T)
 =h(E_A|c_T)+h(E_B|X_B,c_T)$. Hence,
 \begin{align}
 &T_1=h(E_A|X_A,c_T)
 -h(\mathcal{E}_1|X_A,X_B,c_T)\notag\\
 &\,\,+h(E_B|X_B,c_T)=0.
 \end{align}
 Note that $h(\mathcal{E}_1|X_A,X_B,c_T)=h(E_A,E_B|X_A,X_B,c_T)=
 h(E_A|X_A,X_B,c_T)+h(E_B|X_A,X_B,c_T)=h(E_A|X_A,c_T)+h(E_B|X_B,c_T)$. Also note that $\mathcal{E}_1$ consists of the two components $E_A$ and $E_B$ that are functions of the independent $X_A$ and $X_B$ respectively.

The second term in \eqref{eq:IA1B1cTE} is
 \begin{align}\label{eq:IXAYBXBcTE1}
 &T_2=I(X_A;Y_B|c_T,E_A)\notag\\
 &=h(Y_B|c_T,E_A)-h(Y_B|X_A,c_T,E_A)\notag\\
 &=[h(Y_B,E_A|c_T)-h(E_A|c_T)]-[h(Y_B,E_A|X_A,c_T)\notag\\
 &\,\,-h(E_A|X_A,c_T)]\notag\\
 &=[h(Y_B,E_A|c_T)-h(Y_B,E_A|X_A,c_T)]\notag\\
 &\,\,-[h(E_A|c_T)-h(E_A|X_A,c_T)]\notag\\
 &=m_A\mathbb{E}\{\log|p_A\mathbf{g}_A'\mathbf{g}_A'^H+\mathbf{D}_B|\cdot|\mathbf{D}_B^{-1}|\}
 \notag\\
&\,\,-m_A\mathbb{E}\{\log|p_A\mathbf{g}_A\mathbf{g}_A^H/\sigma_{E,A}^2+\mathbf{I}_{n_E}|\}\notag\\
 &=m_A\mathbb{E}\{\log(p_A\mathbf{g}_A'^H\mathbf{D}_B^{-1}\mathbf{g}_A'+1)\}
 \notag\\
 &\,\,-m_A\mathbb{E}\{\log (p_A\|\mathbf{g}_A\|^2/\sigma_{E,A}^2+1)\}\notag\\
 &=m_A\mathbb{E}\left \{\log\left (1+\frac{p_A|h_{B,A}|^2/\sigma_B^2}{p_A\|\mathbf{g}_A\|^2/\sigma_{E,A}^2+1}\right )\right \},
 \end{align}
 where $\mathbf{g}_A'=[h_{B,A},\mathbf{g}_A^T]^T$ and $\mathbf{D}_B=diag(\sigma_B^2,\sigma_{E,B}^2\mathbf{I}_{n_E})$.

 The third term $T_3$ in \eqref{eq:IA1B1cTE} is symmetric with $T_2$ in terms of  ``A'' and ``B''.
 %

 The fourth term in \eqref{eq:IA1B1cTE} is
 \begin{align}
 &T_4=h(Y_A|X_B,X_A,h_T,\mathcal{E}_1)-
 h(Y_A|Y_B,X_B,X_A,h_T,\mathcal{E}_1)\notag\\
 &=h(Y_A,\mathcal{E}_1|X_B,X_A,h_T)-h(\mathcal{E}_1|X_B,X_A,h_T)\notag\\
 &\,\,-[h(Y_A,Y_B,\mathcal{E}_1|X_B,X_A,h_T)-h(Y_B,\mathcal{E}_1|X_B,X_A,h_T)].
 \end{align}
 Then $T_4=[h(Y_A,E_B|X_B,h_T)+h(E_A|X_A,h_T)]
 -[h(E_A|X_A,h_T)+h(E_B|X_B,h_T)]
 -[h(Y_B,E_A|X_A,h_T)+h(Y_A,E_B|X_B,h_T)]
 +
 [h(Y_B,E_A|X_A,h_T)+h(E_B|X_B,h_T)]=0$.


 Therefore, combining the above results into \eqref{eq:CE} yields \eqref{eq:CE3}. The proof of Theorem \ref{Theorem1} is completed.

 \section{Return Channel Transmission After One-Way Probing} \label{sec:return}
 As discussed in section \ref{sec:discussion_theorem1}, the secret-key capacity from one-way probing is always positive and increasing linearly with the number of probing symbols. We will from now on only focus on one-way probing.
 We can now assume $m_A>m_B=0$ without loss of generality.
 In this case, Bob receives $y_B(k)=h_{B,A}x_A(k)+w_B(k)$ for $k=1,\cdots,m_A$ due to channel probing from Alice, and the secret-key capacity is given by $C_B$ in \eqref{eq:CB_one_way}.

 If we follow a generalized SKG procedure shown in \cite{Hua2023}, we need a pre-processing so that both Alice and Bob can obtain good estimates of a common vector.
 Following a similar strategy in \cite{Hua2023}, we now let Bob send out $r(k)=s(k)+y_B(k)$ over a return channel to Alice (and unavoidably another return channel to Eve) where $s(k)$ is a random sequence generated by Bob. We assume that the signals received by Alice and Eve via the return channels are respectively
 \begin{align}\label{}
   &r_A(k)=r(k)+v_A(k)\notag\\
   &=s(k)+h_{B,A}x_A(k)+w_B(k)+v_A(k),
 \end{align}
 \begin{align}\label{}
   &r_E(k)=r(k)+v_E(k)\notag\\
   &=s(k)+h_{B,A}x_A(k)+w_B(k)+v_E(k).
 \end{align}
Here $s(k)$ for $k=1,\cdots,m_A$ are i.i.d. $\mathcal{CN}(0,\sigma_s^2)$, $v_A(k)$ for $k=1,\cdots,m_A$ are i.i.d. $\mathcal{CN}(0,\epsilon_A)$, and $v_E(k)$ for $k=1,\cdots,m_A$ are i.i.d. $\mathcal{CN}(0,\epsilon_E)$. The secret information in $s(k)$ is meant to be received by Alice.

Because of the transmission of $r(k)$ by Bob,  the new secret-key capacity $C_{key}'$ is likely to be different from  $C_{key}=C_B$. To understand $C_{key}'$, we let  the renewed data sets at Alice, Bob and Eve be respectively $\mathcal{A}'=\{\mathcal{A}'_1,h_{A,B}\}=\{X_A,R_A,h_{A,B}\}$, $\mathcal{B}'=\{\mathcal{B}'_1,h_{B,A}\}=\{Y_B,S,h_{B,A}\}$ and $\mathcal{E}' =\{\mathcal{E}'_1,\mathbf{g}_A,\mathbf{g}_B\}= \{E_A,R_E,\mathbf{g}_A,\mathbf{g}_B\}$. Here $S=\{s(k),\forall k\}$, $R_A=\{r_A(k),\forall k\}$ and $R_E=\{r_E(k),\forall k\}$.

There is no need to consider $C_A'$ as it satisfies $C_A'\leq C_A$ (if the return channel is perfect and same for both Alice and Eve) and $C_A$  (with $m_B=0$) is previously shown to be non-positive if Eve's channel from Alice during probing is no weaker than that from Alice to Bob.
 We will next analyze $C_B'=I(\mathcal{A}';\mathcal{B}')-I(\mathcal{B}';\mathcal{E}')$.

\begin{Theorem}\label{Theorem2}
For $\epsilon_A\ll \sigma_B^2$, $\epsilon_E\ll \sigma_B^2$ and $\eta = \frac{\epsilon_E}{\epsilon_B}$, it follows that
  \begin{align}
 &C_{key}'\geq C_B'\geq \alpha' +m_A\xi_{B,A}'+m_A\log \eta
 \end{align}
 with
 \begin{align}\label{}
   &\alpha'=\mathbb{E}\left \{\log\left (\frac{\|\mathbf{x}_A\|^2/(\sigma_s^2+\sigma_B^2)+\frac{1}{1-|\rho|^2}}
   {\|\mathbf{x}_A\|^2/(\sigma_s^2+\sigma_B^2)+1}\right )\right \},
 \end{align}
  \begin{align}\label{eq:xiBAP}
 &\xi_{B,A}'=\mathbb{E}\left \{\log\left ( 1+\frac{\phi_{B,A}\sigma_s^2/\sigma_B^2}{1+\phi_{B,A}+\sigma_s^2/\sigma_B^2}\right ) \right \}.
 \end{align}
Here $\mathbf{x}_A=[x_A(1),\cdots,x_A(m_A)]^T$ and $\phi_{B,A}$ is defined in \eqref{eq:phiB}.
\end{Theorem}

A discussion is shown next before the proof is given.

\subsection{Discussion of Theorem \ref{Theorem2}}\label{sec:discussion_of_theorem2}

  It is clear that $0<\alpha'< \alpha$ where the lower bound is approached when $\|\mathbf{x}_A\|^2/(\sigma_s^2+\sigma_B^2)\gg \frac{1}{1-|\rho|^2}$ which includes the case of a sufficiently large $m_A$, and the upper bound is approached when $\|\mathbf{x}_A\|^2/(\sigma_s^2+\sigma_B^2)\ll 1$ which includes the case of a sufficiently large $\sigma_s^2$.

Also, $\xi_{B,A}'\leq \bar \xi_{B,A}\leq \xi_{B,A}$ with
 \begin{equation}\label{}
   \bar\xi_{B,A}=\mathbb{E}\left \{\log\left ( 1+\eta_s\frac{p_A|h_{B,A}|^2/\sigma_B^2}{
 p_A\|\mathbf{g}_A\|^2/\sigma_{E,A}^2+1}\right ) \right \},
 \end{equation}
 where $\eta_s=\frac{\sigma_s^2/\sigma_B^2}{\sigma_s^2/\sigma_B^2+1}$.
We see that $\xi_{B,A}'= \bar \xi_{B,A}$ is achieved if
 $\sigma_s^2$ is so large that
 \begin{equation}\label{eq:condition}
   (\sigma_s^2/\sigma_B^2+
 1)(p_A\|\mathbf{g}_A\|^2/\sigma_{E,A}^2+1)\gg p_A|h_{B,A}|^2/\sigma_B^2.
 \end{equation}
 And   $\bar \xi_{B,A}= \xi_{B,A}$ if $\sigma_s^2\gg \sigma_B^2$.

 Furthermore, if $\eta\leq 1$ (i.e., Eve's return channel is no noisier than Alice's return channel), $C_B'\leq C_B=C_{key}$.

 For $\eta\leq 1$, there is no need to analyze explicitly $C_E'=I(\mathcal{A}';\mathcal{B}'|\mathcal{E}')$ since the upper bound of secret-key capacity cannot increase after any further processing over the public channels. Namely, we would expect that for $\eta\leq 1$, $C_E'\leq C_E=C_{key}$.

Note that for the case of $\sigma_s^2\to \infty$, $S$ seems completely exposed to Eve. How could any secrecy in $S$ be still protected?
To help answer this question, we will in the next section consider the case where Bob only relies on $S=\{s(k),\forall k\}$ after $R=\{r(k),\forall k\}$ has been sent.

\subsection{Proof of Theorem \ref{Theorem2}}

\subsubsection{Analysis of $I(\mathcal{A}';\mathcal{B}')$}
 It follows that
 \begin{align}\label{eq:IAPBPa}
 &I(\mathcal{A}';\mathcal{B}')=I(h_{A,B};\mathcal{B}')+I(\mathcal{A}'_1;\mathcal{B}'|h_{A,B})\notag\\
 &=I(h_{A,B};h_{B,A})+I(h_{A,B};\mathcal{B}'_1|h_{B,A})\notag\\
 &\,\,+I(\mathcal{A}'_1;h_{B,A}|h_{A,B})+
 I(\mathcal{A}'_1;\mathcal{B}'_1|h_T)
 \end{align}
 Here we know $I(h_{A,B};h_{B,A})=-\log(1-|\rho|^2)$. The 2nd term in \eqref{eq:IAPBPa} is
 \begin{align}
 &I(h_{A,B};\mathcal{B}'_1|h_{B,A})=I(h_{A,B};Y_B|S,h_{B,A})\notag\\
 &=h(Y_B|S,h_{B,A})-h(Y_B|h_{A,B},S,h_{B,A})\notag\\
 &=h(Y_B|h_{B,A})-h(Y_B|h_{B,A})=0.
 \end{align}
 The 3rd term in \eqref{eq:IAPBPa} is
 \begin{align}
 &I(\mathcal{A}'_1;h_{B,A}|h_{A,B})=I(R_A;h_{B,A}|X_A,h_{A,B})\notag\\
 &=h(R_A|X_A,h_{A,B})-h(R_A|X_A,h_{A,B},h_{B,A})\notag\\
 &=h(R_A|X_A,h_{A,B})-h(R_A|X_A,h_{B,A}),
 \end{align}
 where $R_A$, when conditioned on $h_{B,A}$, is independent of $h_{A,B}$. Recall $r_A(k)=s(k)+h_{B,A}x_A(k)+w_B(k)+v_A(k)$, or equivalently
 \begin{equation}\label{}
   \mathbf{r}_A=\mathbf{s}+h_{B,A}\mathbf{x}_A+\mathbf{w}_B+\mathbf{v}_A
 \end{equation}
 with $\mathbf{r}_A\doteq[r_A(1),\cdots,r_A(m_A)]^T$. Note that given $h_{A,B}$, $h_{B,A}$ is Gaussian distributed with the variance $1-|\rho|^2$. So, the PDF of $\mathbf{r}_A$ given $h_{B,A}$ and $X_A$ is $\mathcal{CN}(h_{B,A}\mathbf{x}_A,\mathbf{R}_r)$ with $\mathbf{R}_r=g_A\mathbf{I}_{m_A}$ and $g_A=\sigma_s^2+\sigma_B^2+\epsilon_A$. And the PDF of $\mathbf{r}_A$ given $h_{A,B}$ and $X_A$ is $\mathcal{CN}(\hat h_{B,A}\mathbf{x}_A,\mathbf{R}_r')$ with $\mathbf{R}_r'=(1-|\rho|^2)\mathbf{x}_A\mathbf{x}_A^H+g_A\mathbf{I}_{m_A}$.
Therefore,
  \begin{align}
 &I(\mathcal{A}'_1;h_{B,A}|h_{A,B})=\mathbb{E}\{\log|\mathbf{R}_r'|-\log|\mathbf{R}_r|\}\notag\\
 &=\mathbb{E}\{\log((1-|\rho|^2)\|\mathbf{x}_A\|^2/g_A+1)\},
 \end{align}
 which, unlike the second term in \eqref{eq:IAPBPa}, is not zero.

 For the 4th term in \eqref{eq:IAPBPa}, we have
 \begin{align}
 &I(\mathcal{A}'_1;\mathcal{B}'_1|h_T)=I(X_A;\mathcal{B}'_1|h_T)+I(R_A;\mathcal{B}'_1|X_A,h_T)\notag\\
 &=I(X_A;Y_B|S,h_T)+I(R_A;S|X_A,h_T)\notag\\
 &\,\,+I(R_A;Y_B|S,X_A,h_T),
 \end{align}
 where we have used $I(X_A;S|h_T)=0$. Furthermore,
 \begin{align}\label{}
   &I(X_A;Y_B|S,h_T)=h(Y_B|S,h_T)-h(Y_B|X_A,S,h_T)\notag\\
   &=
   m_A\mathbb{E}\{\log(1+p_A|h_{B,A}|^2/\sigma_B^2)\},
 \end{align}
 \begin{align}\label{}
   &I(R_A;S|X_A,h_T)=h(R_A|X_A,h_T)-h(R_A|S,X_A,h_T)\notag\\
   &=m_A\log(1+\sigma_s^2/(\sigma_B^2+\epsilon_A)),
 \end{align}
 \begin{align}\label{}
  & I(R_A;Y_B|S,X_A,h_T)=h(R_A|S,X_A,h_T)\notag\\
  &\,\,-h(R_A|Y_B,S,X_A,h_T)\notag\\
  &=m_A\log(1+\sigma_B^2/\epsilon_A).
 \end{align}

Using the above results in \eqref{eq:IAPBPa}, subject to $\epsilon_A\ll \sigma_B^2$, yields
\begin{align}\label{eq:IAPBP}
&I(\mathcal{A}';\mathcal{B}')=\mathbb{E}\left \{\log\left (\frac{\|\mathbf{x}_A\|^2}{\sigma_s^2+\sigma_B^2}
+\frac{1}{1-|\rho|^2}\right )\right \}\notag\\
&\,\,+m_A\mathbb{E}\left \{\log(1+p_A|h_{B,A}|^2/\sigma_B^2)\right \}\notag\\
&\,\,+m_A\log(1+\sigma_s^2/\sigma_B^2)
+m_A\log(\sigma_B^2/\epsilon_A).
\end{align}
Note that the last term does not converge as $\epsilon_A\to 0$. We will expect $I(\mathcal{B}';\mathcal{E}')$ to have a similar term to ``balance'' it out.

 \subsubsection{Analysis of $I(\mathcal{B}';\mathcal{E}')$}
 We will use $g_T$ to denote $\{\mathbf{g}_A,\mathbf{g}_B\}$, and $c_T'$ to denote $\{h_{B,A},\mathbf{g}_A,\mathbf{g}_B\}$.
 It follows that
 \begin{align}\label{eq:IBPEP_0}
 &I(\mathcal{B}';\mathcal{E}')=I(h_{B,A};\mathcal{E}')+I(\mathcal{B}'_1;\mathcal{E}'|h_{B,A})\notag\\
 &=I(h_{B,A};\mathcal{E}'_1|g_T)+
 I(\mathcal{B}'_1;\mathcal{E}'_1|c_T').
 \end{align}
 Here,
 \begin{align}
 &I(h_{B,A};\mathcal{E}'_1|g_T)=I(h_{B,A};R_E|E_A,g_T)\notag\\
 &=h(h_{B,A}|E_A,g_T)-h(h_{B,A}|R_E,E_A,g_T).
 \end{align}
  Recall $r_E(k)=s(k)+h_{B,A}x_A(k)+w_B(k)+v_E(k)$, and $\mathbf{e}_A(k)=\mathbf{g}_A x_A(k)+\mathbf{w}_{E,A}$. It follows that $h(h_{B,A}|E_A,g_T)=h(h_{B,A})=\log(e\pi)$. Furthermore,
  \begin{align}\label{}
    &h(h_{B,A}|R_E,E_A,g_T)\geq h(h_{B,A}|R_E,E_A,g_T,W_{E,A})\notag\\
    &    =h(h_{B,A}|R_E,X_A)
  \end{align}
  where $W_{E,A}$ consists of $\mathbf{w}_{E,A}(k)$ for all $k$. We can write
  \begin{equation}\label{}
    \mathbf{r}_E = [r_E(1),\cdots,r_E(m_A)]^T=
    \mathbf{s}+h_{B,A}\mathbf{x}_A+\mathbf{w}_B+\mathbf{v}_E.
  \end{equation}
  Given $\{R_E,X_A\}$ (or equivalently $\{\mathbf{r}_E,\mathbf{x}_A\}$), the PDF of $h_{B,A}$ is $\mathcal{CN}(\hat h_{B,A},\epsilon_h)$ with $\epsilon_h = \frac{g_E}{\|\mathbf{x}_A\|^2+g_E}$ and $g_E=\sigma_s^2+\sigma_B^2+\epsilon_E$. Hence,
 \begin{equation}\label{}
    h(h_{B,A}|R_E,E_A,g_T)\geq \log(e\pi)+\mathbb{E}\{\log\epsilon_h\},
  \end{equation}
  and therefore the 1st term in \eqref{eq:IBPEP_0} is
   \begin{align}\label{eq:IBPEP_0_1}
 &I(h_{B,A};\mathcal{E}'_1|g_T)\leq -\mathbb{E}\{\log\epsilon_h\}\notag\\
 & =\mathbb{E}\{\log (1+\|\mathbf{x}_A\|^2/g_E)\}.
 \end{align}

 For the 2nd term in \eqref{eq:IBPEP_0}, we can replace $c'_T$ by (the simpler notation) $c_T$ without affecting the result. We can further write that 2nd term as
 \begin{align}\label{eq:IBPEP}
 &I(\mathcal{B}_1';\mathcal{E}_1'|c_T)
 =I(S;\mathcal{E}_1'|c_T)+I(Y_B;\mathcal{E}_1'|S,c_T)\notag\\
 &=I(S;E_A|c_T)+I(S;R_E|E_A,c_T)\notag\\
 &\,\,+I(Y_B;R_E|S,c_T)+I(Y_B;E_A|R_E,S,c_T).
 \end{align}
 Here $I(S;E_A|c_T)=0$. Also,
 \begin{align}
 &I(S;R_E|E_A,c_T)=h(R_E|E_A,c_T)-h(R_E|S,E_A,c_T)\notag\\
 &=h(R_E,E_A|c_T)-h(E_A|c_T)\notag\\
 &\,\,-[h(R_E,E_A|S,c_T)-h(E_A|S,c_T)]\notag\\
 &=h(R_E,E_A|c_T)-h(R_E,E_A|S,c_T),
 \end{align}
 \begin{align}
 &I(Y_B;R_E|S,c_T)=h(R_E|S,c_T)-h(R_E|Y_B,S,c_T)\notag\\
 &=h(R_E|S,c_T)-h(R_E,Y_B|S,c_T)+h(Y_B|S,c_T),
 \end{align}
  \begin{align}
 &I(Y_B;E_A|R_E,S,c_T)=h(Y_B|R_E,S,c_T)\notag\\
 &\,\,-h(Y_B|E_A,R_E,S,c_T)\notag\\
 &=h(Y_B,R_E|S,c_T)-h(R_E|S,c_T)\notag\\
 &\,\,-h(Y_B,E_A,R_E|S,c_T)+h(E_A,R_E|S,c_T).
 \end{align}
 Using the above results, and $h(Y_B|S,c_T)=h(Y_B|c_T)$, in \eqref{eq:IBPEP} yields
 \begin{align}\label{eq:IBPEP2}
 &I(\mathcal{B}_1';\mathcal{E}_1'|c_T)=h(R_E,E_A|c_T)
 +h(Y_B|c_T)\notag\\
&\,\, -h(Y_B,E_A,R_E|S,c_T).
 \end{align}

 For $h(R_E,E_A|c_T)$ in \eqref{eq:IBPEP2}, we recall $r_E(k)=s(k)+h_{B,A}x_A(k)+w_B(k)+v_E(k)$ and $\mathbf{e}_A(k)=\mathbf{g}_A x_A(k)+\mathbf{w}_{E,A}(k)$. Also write
 \begin{equation}\label{}
   \mathbf{e}'_A(k)\doteq [r_E(k),\mathbf{e}^T_A(k)]^T=\mathbf{g}_A' x_A(k)+\mathbf{w}'_{E,A}(k)
 \end{equation}
 with $\mathbf{g}_A'=[h_{B,A},\mathbf{g}_A^T]^T$ and $\mathbf{w}'_{E,A}(k)=[s(k)+w_B(k)+v_E(k),\mathbf{w}_{E,A}^T]^T$. The PDF of $\mathbf{e}'_A(k)$ given $c_T$ is $\mathcal{CN}(0,p_A\mathbf{g}_A'\mathbf{g}_A'^H+\mathbf{D}_E')$ with $\mathbf{D}_E'=diag(g_E,\sigma_{E,A}^2\mathbf{I}_{n_E})$ and $g_E=\sigma_s^2+\sigma_B^2+\epsilon_E$. So,
 \begin{align}\label{}
   &h(R_E,E_A|c_T)=m_A(n_E+1)\log(e\pi)\notag\\
   &\,\,+m_A\mathbb{E}\{\log|p_A\mathbf{g}_A'\mathbf{g}_A'^H+\mathbf{D}_E'|\}\notag\\
   &=m_A(n_E+1)\log(e\pi)+m_A\mathbb{E}\{\log (g_E
\sigma_{E,A}^{2n_E}[
p_A|h_{B,A}|^2/g_E\notag\\
&\,\,+p_A\|\mathbf{g}_A\|^2/\sigma_{E,A}^2+1])\}.
 \end{align}

 It is obvious that the 2nd term in \eqref{eq:IBPEP2} is $h(Y_B|c_T)=m_A\log(e\pi)+m_A\mathbb{E}\{\log(p_A|h_{B,A}|^2+\sigma_B^2)$.

 For $h(Y_B,E_A,R_E|S,c_T)$ in \eqref{eq:IBPEP2}, we let
\begin{align}
&\mathbf{y}(k)\doteq\left [\begin{array}{c}
                        r_E(k)-s(k) \\
                        y_B(k) \\
                        \mathbf{e}_A(k)
                      \end{array}
\right ]=\mathbf{g}_A''x_A(k) + \mathbf{1}_2w_B(k) \notag\\
&\,\,+\mathbf{w}_{E,A}''+\mathbf{1}_1v_E(k)
\end{align}
where $\mathbf{g}_A''=[h_{B,A},h_{B,A},\mathbf{g}_A^T]^T$, $\mathbf{1}_2=[1,1,0,\cdots,0]^T$, $\mathbf{w}_{E,A}''=[0,0,\mathbf{w}_{E,A}^T]^T$, and $\mathbf{1}_1=[1,0,\cdots,0]^T$. It follows that the PDF of $\mathbf{y}(k)$ given $\{S,c_T\}$ is $\mathcal{CN}(*,\mathbf{R}_y)$ with
\begin{equation}\label{}
  \mathbf{R}_y = p_A\mathbf{g}_A''\mathbf{g}_A''^H+\sigma_B^2\mathbf{1}_2\mathbf{1}_2^H
  +\sigma_{E,A}^2\mathbf{I}''_{n_E}+\epsilon_E\mathbf{1}_1\mathbf{1}_1^H
\end{equation}
where $\mathbf{I}''_{n_E}=diag(0,0,\mathbf{I}_{n_E})$.
Equivalently,
\begin{equation}\label{}
  \mathbf{R}_y=\left [\begin{array}{cc}
                        \mathbf{R}_{1,1} & \mathbf{R}_{1,2} \\
                        \mathbf{R}_{1,2}^H & \mathbf{R}_{2,2}
                      \end{array}
   \right ]
\end{equation}
where
\begin{equation}\label{}
  \mathbf{R}_{1,1}=\left [\begin{array}{cc}
                            g_1+\epsilon_E & g_1 \\
                            g_1 & g_1
                          \end{array}
   \right ],
\end{equation}
\begin{equation}\label{}
  \mathbf{R}_{1,2}=p_Ah_{B,A}\mathbf{1}\mathbf{g}_A^H,
\end{equation}
\begin{equation}\label{}
  \mathbf{R}_{2,2}=p_A\mathbf{g}_A\mathbf{g}_A^H+\sigma_{E,A}^2\mathbf{I}_{n_E}
\end{equation}
with $g_1=p_A|h_{B,A}|^2+\sigma_B^2$. Then,
\begin{align}
&|\mathbf{R}_y|=|\mathbf{R}_{1,1}|\cdot |\mathbf{R}_{2,2}-\mathbf{R}_{1,2}^H
\mathbf{R}_{1,1}^{-1}\mathbf{R}_{1,2}|\notag\\
&=g_1\epsilon_E \left |p_A\mathbf{g}_A\mathbf{g}_A^H+\sigma_{E,A}^2\mathbf{I}_{n_E}\right . \notag\\
&\,\,\left . -
p_A^2\mathbf{g}_A|h_{B,A}|^2\mathbf{1}^T\frac{1}{g_1\epsilon_E}\left [ \begin{array}{cc}
                                         g_1 & -g_1 \\
                                         -g_1 & g_1+\epsilon_E
                                       \end{array}
\right ]\mathbf{1}\mathbf{g}_A^H\right |\notag\\
&=g_1\epsilon_E \left |g_2\mathbf{g}_A\mathbf{g}_A^H+\sigma_{E,A}^2\mathbf{I}_{n_E}\right |,
\end{align}
with $g_2 = p_A-\frac{|h_{B,A}|^2p_A^2}{g_1}=\frac{p_A\sigma_B^2}{g_1}$. Finally,
\begin{equation}\label{}
  |\mathbf{R}_y|=g_1\epsilon_E\sigma_{E,A}^{2n_E}(g_2\|\mathbf{g}_A\|^2/\sigma_{E,A}^2+1).
\end{equation}
Note that
\begin{equation}\label{}
  h(Y_B,E_A,R_E|S,c_T)=
  (n_E+2)m_A\log(e\pi)+\mathbb{E}\{\log|\mathbf{R}_y|\}.
\end{equation}

Now combining the above results in \eqref{eq:IBPEP2}, subject to $\epsilon_E\ll \sigma_B^2$, yields
\begin{align}\label{eq:IBPEP3}
 &I(\mathcal{B}_1';\mathcal{E}_1'|c_T)=
 m_A\mathbb{E}\{\log(g_E[
p_A|h_{B,A}|^2/g_E\notag\\
&\,\,+p_A\|\mathbf{g}_A\|^2/\sigma_{E,A}^2+1])\}\notag\\
&\,\,-m_A\mathbb{E}\{\log (\epsilon_E(g_2\|\mathbf{g}_A\|^2/\sigma_{E,A}^2+1))\}.
 \end{align}
Here we notice the term $-m_A\log\epsilon_E$ which would not converge if $\epsilon_E\to 0$. But this term will ``balance'' out the term  $-m_A\log\epsilon_A$ in $I(\mathcal{A}';\mathcal{B}')$.

 \subsubsection{Final form of $C_B'$}
 For $\epsilon_A\ll \sigma_B^2$, $\epsilon_B\ll \sigma_B^2$, and $\frac{\epsilon_E}{\epsilon_A}= \eta$, it is straightforward to verify from \eqref{eq:IAPBP}, \eqref{eq:IBPEP_0}, \eqref{eq:IBPEP_0_1} and \eqref{eq:IBPEP3}  that
 \begin{align}
 &C_B'=I(\mathcal{A}';\mathcal{B}')-I(\mathcal{B}';\mathcal{E}')\notag\\
 &\geq \alpha' +m_A\xi_{B,A}'+m_A\log \eta,
 \end{align}
 where
 \begin{align}\label{}
   &\alpha'=\mathbb{E}\left \{\log\left (\frac{\|\mathbf{x}_A\|^2/(\sigma_s^2+\sigma_B^2)+\frac{1}{1-|\rho|^2}}
   {\|\mathbf{x}_A\|^2/(\sigma_s^2+\sigma_B^2)+1}\right )\right \}\notag\\
   &=\mathbb{E}\left \{\log\left (1+\frac{|\rho|^2}
   {(1-|\rho|^2)(\|\mathbf{x}_A\|^2/(\sigma_s^2+\sigma_B^2)+1)}\right )\right \},
 \end{align}
 \begin{align}
 &\xi_{B,A}'
 =\mathbb{E}\left \{\log\left ( 1+\frac{(\sigma_s^2/\sigma_B^2)\phi_{B,A}}{\phi_{B,A}+
 \sigma_s^2/\sigma_B^2+
 1}\right ) \right \},
 \end{align}
 and $\phi_{B,A}$ is defined in \eqref{eq:phiB}.
The proof of Theorem \ref{Theorem2} is completed.

\section{Secret-Key Capacity Based on Pruned Set at Bob}\label{sec:pruned}

An intriguing question from section \ref{sec:discussion_of_theorem2} is: how can $\sigma_s^2\to \infty$ be the optimal choice? Is it because the term $h_{B,A}x_A(k)$ in $h_{B,A}x_A(k)+s(k)$ (equivalently $Y_B$ in $S+Y_B$) is still hidden from Eve when $\sigma_s^2\to \infty$? To answer this question, we now restrict the data set to be used by Bob (after he has sent out $R=S+Y_B$) to be $\mathcal{B}''=\{S,h_{B,A}\}$ which is $\mathcal{B}'$ without $Y_B$. But Alice and Eve may still use  respectively $\mathcal{A}'=\{X_A,R_A,h_{A,B}\}$ and $\mathcal{E}'=\{E_A,R_E,\mathbf{g}_A\}$.

Note that since we have chosen $m_B=0$ and $m_A>0$, then there is no $E_B$, and $\mathbf{g}_B$ is useless for Eve (subject to independence between $\{h_{A,B},h_{B,A}\}$ and $\{\mathbf{g}_A,\mathbf{g}_B\}$).

We will consider $C_B''=I(\mathcal{A}';\mathcal{B}'')-I(\mathcal{B}'';\mathcal{E}')$.

\begin{Theorem}\label{Theorem3}
For $\epsilon_A\ll \sigma_B^2$ and $\epsilon_E\ll \sigma_B^2$,
  \begin{align}
&C_B''\geq \alpha' +m_A\xi_{B,A}',
\end{align}
where $\alpha'$ and $\xi_{B,A}'$ are the same as those in Theorem \ref{Theorem2}.
\end{Theorem}

\subsection{Discussion of Theorem \ref{Theorem3}}
Unlike $C_B'$, $C_B''$ is invariant to $\eta$ subject to $\epsilon_A\ll \sigma_B^2$ and $\epsilon_E\ll \sigma_B^2$. Given the previous discussions of $\alpha'$ and $\xi_{B,A}'$, we have $C_B''\leq C_B=C_{key}$ where the equality is approached if $\sigma_s^2$ is sufficiently large.

Once again, we see that the optimal secret-key capacity is achieved by $\sigma_s^2\to\infty$. One explanation now is that although $S$ under $\sigma_s^2\to\infty$ is almost fully exposed to Eve over the return channel, Alice can always get a better estimate of $S$ due to her knowledge of $X_A$. In the next section, we will provide such a comparison analytically.

Also note that as discussed in section \ref{sec:discussion_of_theorem2}, for a large $m_A$, $\alpha'$ diminishes. In other words, as $m_A$ increases, the secrecy contributed by the correlation between $h_{A,B}$ and $h_{B,A}$ becomes less and less significant.

\subsection{Proof of Theorem \ref{Theorem3}}

\subsubsection{Analysis of $I(\mathcal{A}';\mathcal{B}'')$}
We know
\begin{align}\label{eq:IAPBPP}
&I(\mathcal{A}';\mathcal{B}'')=I(h_{A,B};\mathcal{B}'')+I(\mathcal{A}'_1;\mathcal{B}''|h_{A,B})\notag\\
&=I(h_{A,B};h_{B,A})+I(h_{A,B};S|h_{B,A})
+I(\mathcal{A}'_1;h_{B,A}|h_{A,B})\notag\\
&\,\,+I(\mathcal{A}'_1;S|h_{B,A},h_{A,B}),
\end{align}
where the first two terms are given by $I(h_{A,B};h_{B,A})=-\log(1-|\rho|^2)$ and $I(h_{A,B};S|h_{B,A})=0$.

For the last two terms in \eqref{eq:IAPBPP}, recall $r_A(k)=s(k)+h_{B,A}x_A(k)+w_B(k)+v_A(k)$, or equivalently,
\begin{equation}\label{}
  \mathbf{r}_A=\mathbf{s}+h_{B,A}\mathbf{x}_A+\mathbf{w}_B+\mathbf{v}_A(k).
\end{equation}

Then, the 3rd term in \eqref{eq:IAPBPP} is
\begin{align}
&I(\mathcal{A}'_1;h_{B,A}|h_{A,B})
=I(R_A;h_{B,A}|X_A,h_{A,B})\notag\\
&=h(R_A|X_A,h_{A,B})-h(R_A|h_{B,A},X_A,h_{A,B})\notag\\
&=\mathbb{E}\{\log|\mathbf{R}_1|-\log|\mathbf{R}_2|\}\notag\\
&=\mathbb{E}\left \{\log\left (\frac{1-|\rho|^2}{\sigma_s^2+\sigma_B^2+\epsilon_A}\|\mathbf{x}_A\|^2+1\right )\right \},
\end{align}
where $\mathbf{R}_1=(1-|\rho|^2)\mathbf{x}_A\mathbf{x}_A^H+
  (\sigma_s^2+\sigma_B^2+\epsilon_A)\mathbf{I}_{m_A}$, and $\mathbf{R}_2=
  (\sigma_s^2+\sigma_B^2+\epsilon_A)\mathbf{I}_{m_A}$.

The 4th term in \eqref{eq:IAPBPP} is
\begin{align}
&I(\mathcal{A}'_1;S|h_{B,A},h_{A,B})=I(R_A;S|X_A,h_{B,A},h_{A,B})\notag\\
&=h(R_A|X_A,h_T)-h(R_A|S,X_A,h_T)\notag\\
&=\log|(\sigma_s^2+\sigma_B^2+\epsilon_A)\mathbf{I}_{m_A}|
-\log|(\sigma_B^2+\epsilon_A)\mathbf{I}_{m_A}|\notag\\
&=m_A\log\left (1+\frac{\sigma_s^2}{\sigma_B^2+\epsilon_A}\right ).
\end{align}

Combining the above results in \eqref{eq:IAPBPP} yields
\begin{align}\label{eq:IAPBPP2}
&I(\mathcal{A}';\mathcal{B}'')=-\log(1-|\rho|^2)\notag\\
&\,\,+
\mathbb{E}\left \{\log\left (\frac{1-|\rho|^2}{\sigma_s^2+\sigma_B^2+\epsilon_A}\|\mathbf{x}_A\|^2+1\right )\right \}\notag\\
&\,\,+m_A\log\left (1+\frac{\sigma_s^2}{\sigma_B^2+\epsilon_A}\right )\notag\\
&=\mathbb{E}\left \{\log\left (\frac{\|\mathbf{x}_A\|^2}{\sigma_s^2+\sigma_B^2+\epsilon_A}+\frac{1}{1-|\rho|^2}\right )\right \}\notag\\
&\,\,+m_A\log\left (1+\frac{\sigma_s^2}{\sigma_B^2+\epsilon_A}\right ).
\end{align}

\subsubsection{Analysis of $I(\mathcal{B}'';\mathcal{E}')$}
It follows that
\begin{align}\label{eq:IBPPEP}
&I(\mathcal{B}'';\mathcal{E}') = I(h_{B,A};\mathcal{E}')+ I(S;\mathcal{E}'|h_{B,A})\notag\\
&=I(h_{B,A};\mathcal{E}'_1|\mathbf{g}_A)+I(S;\mathcal{E}'_1|h_{B,A},\mathbf{g}_A).
\end{align}
where, as shown in \eqref{eq:IBPEP_0_1},
$I(h_{B,A};\mathcal{E}'_1|\mathbf{g}_A)
  \leq\mathbb{E}\{\log (1+\|\mathbf{x}_A\|^2/g_E)\}$.
Furthermore, the 2nd term in \eqref{eq:IBPPEP} is
\begin{align}
&I(S;\mathcal{E}'_1|h_{B,A},\mathbf{g}_A)
=h(E_A,R_E|h_{B,A},\mathbf{g}_A)\notag\\
&\,\,-h(E_A,R_E|S,h_{B,A},\mathbf{g}_A).
\end{align}
Recall $\mathbf{e}_A(k)=\mathbf{g}_Ax_A(k)+\mathbf{w}_{E,A}(k)$ and $r_E(k) = s(k)+h_{B,A}x_A(k)+w_B(k)+v_E(k)$. Let
\begin{equation}\label{}
  \mathbf{e}_A'(k)=[r_E(k),\mathbf{e}_A^T(k)]^T
  =\mathbf{g}_A'x_A(k)+\mathbf{w}_{E,A}'(k)
\end{equation}
where $\mathbf{g}_A'=[h_{B,A},\mathbf{g}_A^T]^T$,  and $\mathbf{w}_{E,A}'(k)=[s(k)+w_B(k)+v_E(k),\mathbf{w}_{E,A}^T]^T$. Recall $g_E=\sigma_s^2+\sigma_B^2+\epsilon_E$. Also let $g'_E=\sigma_B^2+\epsilon_E$. It follows that
\begin{align}
&I(S;\mathcal{E}'_1|h_{B,A},\mathbf{g}_A) = m_A\mathbb{E}\{\log|\mathbf{R}_{1,E}|-\log|\mathbf{R}_{2,E}|\}
\end{align}
with
\begin{align}\label{}
  &|\mathbf{R}_{1,E}|=|p_A\mathbf{g}_A'\mathbf{g}_A'^H+
  diag(g_E,\sigma_{E,A}^2
  \mathbf{I}_{n_E})|\notag\\
  &=g_E\sigma_{E,A}^{2n_E}(p_A|h_{B,A}|^2/g_E
  +p_A\|\mathbf{g}_A\|^2/\sigma_{E,A}^2+1),
\end{align}
\begin{align}\label{}
  &|\mathbf{R}_{2,E}|=|p_A\mathbf{g}_A'\mathbf{g}_A'^H+diag(g'_E,\sigma_{E,A}^2
  \mathbf{I}_{n_E})|\notag\\
  &=g'_E\sigma_{E,A}^{2n_E}(p_A|h_{B,A}|^2/g'_E
  +p_A\|\mathbf{g}_A\|^2/\sigma_{E,A}^2+1).
\end{align}
Hence, \eqref{eq:IBPPEP} becomes
\begin{align}
&I(\mathcal{B}'';\mathcal{E}') \leq \mathbb{E}\{\log (1+\|\mathbf{x}_A\|^2/g_E)\}\notag\\
&\,\,+ m_A\mathbb{E}\left \{\log \left (\frac{g_E}{g'_E}\right . \right .\notag\\
&\,\,\left .\left .\cdot\frac{p_A|h_{B,A}|^2/g_E
  +p_A\|\mathbf{g}_A\|^2/\sigma_{E,A}^2+1}{p_A|h_{B,A}|^2/g'_E
  +p_A\|\mathbf{g}_A\|^2/\sigma_{E,A}^2+1}\right )\right \}.
\end{align}

\subsubsection{Final form of $C_B''$}
Assume $\sigma_B^2 \gg \epsilon_E$ and $\sigma_B^2 \gg \epsilon_A$. Then, one can verify that
\begin{align}
&C_B''=I(\mathcal{A}';\mathcal{B}'')-I(\mathcal{B}'';\mathcal{E}')\notag\\
&\geq \alpha' +m_A\xi_{B,A}',
\end{align}
where $\alpha'$ and $\xi_{B,A}'$ are the same as in $C_B'$.
The proof of Theorem \ref{Theorem3} is completed.

\section{Optimal Estimation of $S$}\label{sec:estimation_of_s}

In this section, we show that for a large $m_A$, the optimal estimation of $S$ by Alice is always better than that by Eve even if $h_{B,A}$ is known to Eve but unknown to Alice.
Note that once Alice has a good estimate of $S$, both Alice and Bob can follow a standard procedure for SKG, i.e., quantization, reconciliation, and privacy amplification \cite{Huth2016}.

But in the next two sections (i.e., sections \ref{sec:STEEP_analog} and \ref{sec:STEEP_digital}), we will show a new way of looking at the return signals from Bob, which allows the application of WTC based methods (over effective return channels from Bob to Alice and Eve) to achieve the optimal secrecy.

\subsection{Optimal Estimation of $S$ at Alice}

Recall that after Bob sends the return signal $R=S+Y_B$ over a return channel, Alice has $\mathcal{A}'=\{R_A,X_A,h_{A,B}\}$. Also we can write
\begin{equation}\label{}
  \mathbf{r}_A\doteq[r_A(1),\cdots,r_A(m_A)]^T=\mathbf{s}+h_{B,A}\mathbf{x}_A+
  \mathbf{w}_B+\mathbf{v}_A.
\end{equation}
Note that $\mathbf{r}_A$ conditional on $h_{A,B}$ and $\mathbf{x}_A$ is Gaussian while the unconditional PDF of $\mathbf{r}_A$ is not.

Let $\hat h_{B,A}=\rho^*h_{A,B}$ and $\mathbf{r}_A'=\mathbf{r}_A-\hat h_{B,A}\mathbf{x}_A=\mathbf{s}+\Delta h_{B,A}\mathbf{x}_A+
  \mathbf{w}_B+\mathbf{v}_A$. The PDF of $\mathbf{r}_A'$, conditioned on $h_{A,B}$ and $\mathbf{x}_A$, is $\mathcal{CN}(0,\mathbf{R}_r')$ with $\mathbf{R}_r'=(1-|\rho|^2)\mathbf{x}_A\mathbf{x}_A^H+g_A\mathbf{I}_{m_A}$ and $g_A=\sigma_s^2+\sigma_B^2+\epsilon_A$.

The minimum-mean-squared-error (MMSE) estimate of $\mathbf{s}$ by Alice from $\{\mathbf{r}_A,\mathbf{x}_A,h_{A,B}\}$ is
\begin{align}\label{}
  &\mathbf{\hat s}_A =\mathbb{E}\{\mathbf{s}|\mathbf{r}_A,\mathbf{x}_A,h_{A,B}\}\notag\\
  &= \mathbb{E}\{\mathbf{s}{\mathbf{r}'_A}^H|\mathbf{x}_A,h_{A,B}\}
  (\mathbb{E}\{\mathbf{r}'_A{\mathbf{r}'_A}^H|\mathbf{x}_A,h_{A,B}\})^{-1}\mathbf{r}'_A.
\end{align}
Here
\begin{equation}\label{}
 \mathbb{E}\{\mathbf{s}\mathbf{r}_A^H|\mathbf{x}_A,h_{A,B}\}=\sigma_s^2\mathbf{I}_{m_A},
\end{equation}
\begin{equation}\label{}
 \mathbb{E}\{\mathbf{r}'_A{\mathbf{r}'_A}^H|\mathbf{x}_A,h_{A,B}\}
=(1-|\rho|^2)\mathbf{x}_A\mathbf{x}_A^H+g_A\mathbf{I}_{m_A}
\end{equation}
and $g_A=\sigma_s^2+\sigma_B^2+\epsilon_A$. Hence,
\begin{align}\label{}
  &\mathbf{\hat s}_A =\sigma_s^2((1-|\rho|^2)\mathbf{x}_A\mathbf{x}_A^H+g_A\mathbf{I}_{m_A})^{-1}\mathbf{r}'_A
  \notag\\
  &=\frac{\sigma_s^2}{g_A}\left (\mathbf{I}_{m_A}-\frac{(1-|\rho|^2)/g_A}{(1-|\rho|^2)\|\mathbf{x}_A\|^2/g_A+1}
  \mathbf{x}_A\mathbf{x}_A^H\right )\mathbf{r}'_A.
\end{align}

The covariance matrix of $\mathbf{s}$ conditional on $\{\mathbf{r}_A,\mathbf{x}_A,h_{A,B}\}$ is the MSE matrix of $\mathbf{\hat s}$ conditional on $\{\mathbf{r}_A,\mathbf{x}_A,h_{A,B}\}$, which is
\begin{align}
&\mathbf{R}_{\Delta s, A|x}=\sigma_s^2\mathbf{I}_{m_A}
-\sigma_s^4[(1-|\rho|^2)\mathbf{x}_A\mathbf{x}_A^H+g_A\mathbf{I}_{m_A}]^{-1}\notag\\
&=\sigma_s^2\left ((1-\sigma_s^2/g_A)\mathbf{I}_{m_A}
+\frac{\frac{\sigma_s^2}{g_A}\frac{(1-|\rho|^2)}{g_A}}{1+\frac{1-|\rho|^2}{g_A}\|\mathbf{x}_A\|^2}
\mathbf{x}_A\mathbf{x}_A^H\right ).
\end{align}

Since the entries of $\mathbf{x}_A$ are i.i.d. $\mathcal{CN}(0,p_A)$, we can write
\begin{align}
&\mathbb{E}\{\mathbf{R}_{\Delta s, A|x}\}=\sigma_s^2\mathbb{E}\left \{(1-\sigma_s^2/g_A)
\right . \notag\\
&\,\,\left. +\frac{\frac{\sigma_s^2}{g_A}\frac{(1-|\rho|^2)}{g_A}}{1+\frac{1-|\rho|^2}{g_A}\|\mathbf{x}_A\|^2}
\frac{\|\mathbf{x}_A\|^2}{m_A}\right \}\mathbf{I}_{m_A}.
\end{align}
If $\frac{1-|\rho|^2}{g_A}\|\mathbf{x}_A\|^2\gg 1$ and $\sigma_B^2\gg \epsilon_A$,
\begin{align}
&\mathbb{E}\{\mathbf{R}_{\Delta s, A|x}\}=\sigma_s^2\left ((1-\sigma_s^2/g_A)
+\frac{\sigma_s^2/g_A}{m_A}\right )\mathbf{I}_{m_A}\notag\\
&=\sigma_s^2\left (\frac{1}{\sigma_s^2/\sigma_B^2+1}+\frac{1}{m_A}\frac{\sigma_s^2/\sigma_B^2}
{\sigma_s^2/\sigma_B^2+1}\right )\mathbf{I}_{m_A}.
\end{align}
We see that for a large $m_A$, i.e., $m_A\gg \sigma_s^2/\sigma_B^2$,
\begin{equation}\label{eq:lim}
  \mathbb{E}\{\mathbf{R}_{\Delta s, A|x}\}=\bar r_{\Delta s, A|x}\mathbf{I}_{m_A}
\end{equation}
where
$
  \bar r_{\Delta s, A|x}=\sigma_s^2/(\sigma_s^2/\sigma_B^2+1)
$.

\subsection{Optimal Estimation of $S$ at Eve}
Recall $\mathcal{E}'=\{E_A,R_E,\mathbf{g}_A\}$.  Also recall
\begin{equation}\label{}
  \mathbf{e}_A(k)=\mathbf{g}_Ax_A(k)+\mathbf{w}_{E,A}(k)
\end{equation}
and $r_E(k)=s(k)+h_{B,A}x_A(k)+w_B(k)+v_E(k)$. Now we assume $\sigma_B^2\gg \epsilon_E$, and write $r_E(k)=s(k)+h_{B,A}x_A(k)+w_B(k)$. Equivalently, by stacking up all $r_E(k)$ into a vector $\mathbf{r}_E$, we can write
\begin{equation}\label{}
  \mathbf{r}_E = \mathbf{s}+h_{B,A}\mathbf{x}_A+\mathbf{w}_B.
\end{equation}

The MMSE estimate of $x_A(k)$ for all $k$ from $\mathbf{e}_A(k)$ for all $k$ by Eve is
\begin{align}\label{}
  &\hat x_A(k) = p_A\mathbf{g}_A^H(p_A\mathbf{g}_A\mathbf{g}_A^H+\sigma_{E,A}^2\mathbf{I}_{n_E})^{-1}\mathbf{e}_A(k)
  \notag\\
  &=p_A(p_A\|\mathbf{g}_A\|^2+\sigma_{E,A}^2)^{-1}\mathbf{g}_A^H\mathbf{e}_A(k),
\end{align}
and the MSE of $\hat x_A(k)$ is
\begin{align}\label{eq:rDxAE}
  &r_{\Delta x_A}=p_A-p_A^2\mathbf{g}_A^H(p_A\mathbf{g}_A\mathbf{g}_A^H+\sigma_{E,A}^2\mathbf{I}_{n_E})^{-1}
  \mathbf{g}_A\notag\\
  &=p_A-p_A^2(p_A\|\mathbf{g}_A\|^2+\sigma_{E,A}^2)^{-1}\|\mathbf{g}_A\|^2\notag\\
  &=\frac{p_A\sigma_{E,A}^2}{p_A\|\mathbf{g}_A\|^2+\sigma_{E,A}^2}\notag\\
  &=\frac{p_A}{p_A\|\mathbf{g}_A\|^2/\sigma_{E,A}^2+1}.
\end{align}

Equivalently, the MMSE estimate of $\mathbf{x}_A^T=[x_A(1),\cdots,x_A(m_A)]$ from $\mathbf{E}_A=[\mathbf{e}_A(1),\cdots,\mathbf{e}_A(m_A)]$ is
\begin{align}\label{}
  &\mathbf{\hat x}_A^T = p_A(p_A\|\mathbf{g}_A\|^2+\sigma_{E,A}^2)^{-1}\mathbf{g}_A^H\mathbf{E}_A,
\end{align}
and the MSE matrix of $\mathbf{\hat x}_A$ is
\begin{equation}\label{}
  \mathbf{R}_{\Delta x_A}=r_{\Delta x_A}\mathbf{I}_{m_A}.
\end{equation}

If Eve also knows $h_{B,A}$, the MMSE estimate of $\mathbf{s}$ from $\{\mathbf{r}_E,\mathbf{E}_A,\mathbf{g}_A,h_{B,A}\}$ by Eve is
\begin{align}
&\mathbf{\hat s}_E = \sigma_s^2(\sigma_s^2+|h_{B,A}|^2r_{\Delta x_A}+\sigma_B^2)^{-1}(\mathbf{r}_E-h_{B,A}\mathbf{\hat x}_A),
\end{align}
and the MSE matrix of $\mathbf{\hat s}_E $ is
\begin{align}
&\mathbf{R}_{\Delta s_E}=r_{\Delta  s, E}\mathbf{I}_{m_A}
\end{align}
with
\begin{align}\label{}
  &r_{\Delta  s, E}=\sigma_s^2(1-\sigma_s^2(\sigma_s^2+|h_{B,A}|^2r_{\Delta x_A}+\sigma_B^2)^{-1})\notag\\
  &=\sigma_s^2\frac{|h_{B,A}|^2r_{\Delta x_A}+\sigma_B^2}{\sigma_s^2+|h_{B,A}|^2r_{\Delta x_A}+\sigma_B^2}\notag\\
  &=\sigma_s^2\frac{r_{\Delta x_A}|h_{B,A}|^2/\sigma_B^2+1}{\sigma_s^2/\sigma_B^2+r_{\Delta x_A}|h_{B,A}|^2/\sigma_B^2+1}\notag\\
  &=\sigma_s^2\frac{\frac{p_A|h_{B,A}|^2/\sigma_B^2}{p_A\|\mathbf{g}\|^2/\sigma_{E,A}^2+1}+1}
  {\sigma_s^2/\sigma_B^2+\frac{p_A|h_{B,A}|^2/\sigma_B^2}{p_A\|\mathbf{g}\|^2/\sigma_{E,A}^2+1}+1}
  \notag\\
  &>\sigma_s^2\frac{1}{\sigma_s^2/\sigma_B^2+1}=\bar r_{\Delta  s, A}.
\end{align}

Recall \eqref{eq:phiB}. Then,
\begin{align}\label{}
  &r_{\Delta  s, E}
  =\sigma_s^2\frac{\phi_{B,A}+1}
  {\sigma_s^2/\sigma_B^2+\phi_{B,A}+1},
\end{align}
which increases with $\phi_{B,A}$.

It is clear that
\begin{equation}\label{}
\phi_{B,A}\leq \frac{|h_{B,A}|^2/\sigma_B^2}{\|\mathbf{g}\|^2/\sigma_{E,A}^2}=\bar \phi_{B,A},
\end{equation}
where the equality is approached if  $p_A\|\mathbf{g}\|^2/\sigma_{E,A}^2\gg 1$. The upper bound of $\phi_{B,A}$ is a ratio of the strength of users' probing channel over that of Eve's probing channel. If users' probing channel and Eve's probing channel have an identical and high SNR, i.e., $p_A|h_{B,A}|^2/\sigma_B^2=p_A\|\mathbf{g}\|^2/\sigma_{E,A}^2\gg 1$, then
$
\bar \phi_{B,A}=1
$.
If Eve's probing channel has a low SNR, i.e., $p_A\|\mathbf{g}\|^2/\sigma_{E,A}^2\ll 1$, then
$
\phi_{B,A}=p_A|h_{B,A}|^2/\sigma_B^2
$.
In this case, if the users' probing channel also has a low SNR, then
$
\phi_{B,A}=p_A|h_{B,A}|^2/\sigma_B^2\ll 1
$.

\subsection{Comparison of MSEs of estimated $S$}

Let $\eta_{A,E}=\frac{\bar r_{\Delta  s, A}}{r_{\Delta  s, E}}$ which is the ratio of the MSE of the estimated $S$ at Alice over that at Eve subject to $m_A\gg\sigma_s^2/\sigma_B^2$ and $h_{B,A}$ being known to Eve but unknown to Alice. Then, it follows that
\begin{align}\label{}
  &\eta_{A,E}
  =\frac{\sigma_s^2/\sigma_B^2+\phi_{B,A}+1}{(\sigma_s^2/\sigma_B^2+1)(\phi_{B,A}+1)},
\end{align}
which is a decreasing function of $\sigma_s^2$. Furthermore,
\begin{align}\label{}
  &\frac{1}{\phi_{B,A}+1}<\eta_{A,E}<1,
\end{align}
where the lower bound is approached if
 $\sigma_s^2/\sigma_B^2\gg\phi_{B,A}+1$, and the upper bound (which is undesirable) is approached if $\sigma_s^2\to\ 0$.

 We see that the optimal estimation of $S$ by Alice is always better than that by Eve, and the gap is maximized (i.e., $\eta_{A,E}$ is minimized) when $\sigma_s^2\to \infty$.

 Example 1: If $p_A|h_{B,A}|^2/\sigma_B^2=p_A\|\mathbf{g}\|^2/\sigma_{E,A}^2\gg 1$, then $\phi_{B,A}=1$. If in addition $\sigma_s^2/\sigma_B^2\gg 2$, then $\eta_{A,E}=\frac{1}{2}$.

 Example 2: If $p_A|h_{B,A}|^2/\sigma_B^2=\frac{1}{2}p_A\|\mathbf{g}\|^2/\sigma_{E,A}^2\gg 1$, then $\phi_{B,A}=\frac{1}{2}$. If in addition $\sigma_s^2/\sigma_B^2\gg 1.5$, then $\eta_{A,E}=\frac{2}{3}$.

 In both examples,
 $\frac{\bar r_{\Delta s, A|x}}{\sigma_s^2}=(1-\sigma_s^2/g_A)=\frac{\sigma_B^2}{\sigma_s^2}
$.

\section{STEEP for Analog Channels}\label{sec:STEEP_analog}

\begin{figure}
  \centering
  \includegraphics[width=2.5in]{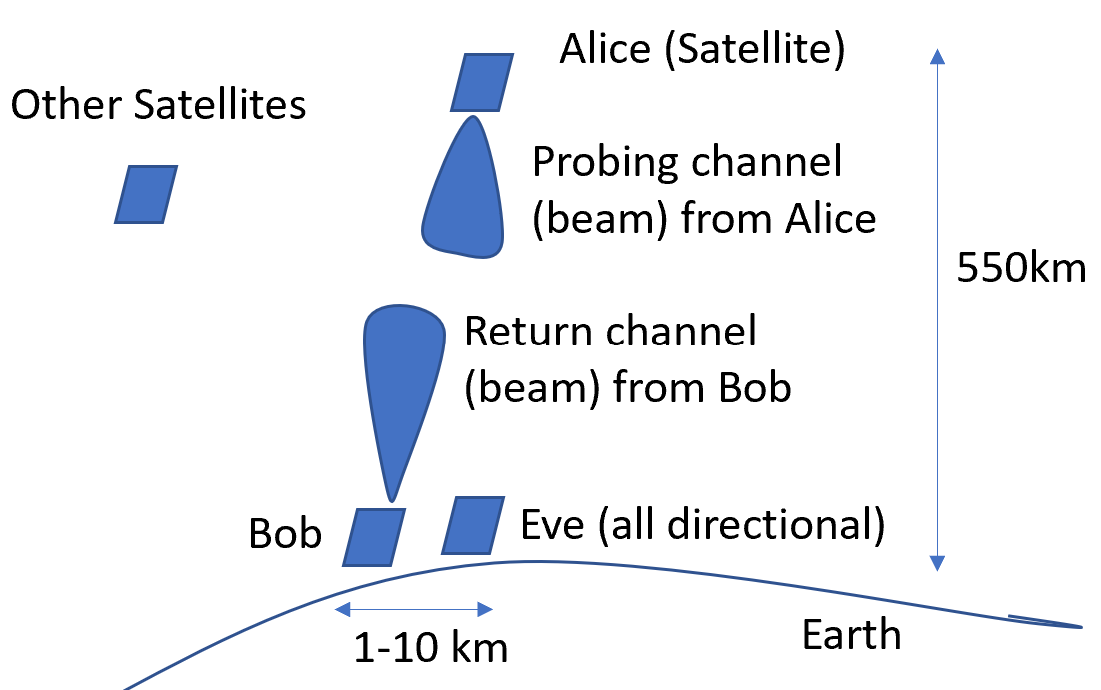}
  \caption{A potential application of STEEP where the channel strength for Eve during probing is either weaker or stronger than that for Bob.}\label{fig:satellite}
\end{figure}

We now combine the previous insights to formulate a scheme we call ``secret-message transmission by echoing encrypted probes (STEEP)'' for analog probing and return channels. A potential application of STEEP is illustrated in Fig. \ref{fig:satellite} where Eve's channel during probing is allowed to be stronger than Bob's channel.

Recall that after channel probing,  Bob and Eve receive respectively
\begin{equation}\label{}
  y_B(k)=h_{B,A}x_A(k)+w_B(k),
\end{equation}
\begin{equation}\label{}
  \mathbf{e}_A(k)=\mathbf{g}_Ax_A(k)+\mathbf{w}_{E,A}(k),
\end{equation}
with $k=1,\cdots,m_A$.

We will now assume that Bob has sent the complete information of $h_{B,A}$ to Alice (via feedback) so that $h_{B,A}$ is known to both Alice and Eve. But $\mathbf{g}_A$ is only known to Eve.

Then Bob sends out $r(k)=y_B(k)+s(k)$ over return channels so that Alice and Eve receive respectively
\begin{align}\label{}
  &y_{A,B}(k)=r(k)+v_A(k)\notag\\
  &=s(k)+h_{B,A}x_A(k)+w_B(k)+v_A(k),
  \end{align}
  \begin{align}\label{}
  &y_{E,B}(k)=r(k)+v_E(k)\notag\\
  &=s(k)+h_{B,A}x_A(k)+w_B(k)+v_E(k).
\end{align}
We will assume that $\sigma_B^2\gg \epsilon_A$ and $\sigma_B^2\gg \epsilon_E$ so that $v_A(k)$ and $v_E(k)$ in the above can be dropped.

\subsection{Effective return channel from Bob to Alice}

We can write the MMSE estimate of $s(k)$ from $\{x_A(k),y_{A,B}(k),h_{B,A}, \forall k\}$ as
\begin{align}\label{}
  &\hat s_A(k) = \mathbb{E}\{s(k)|x_A(k),y_{A,B}(k),h_{B,A}\}\notag\\
  &=\mathbb{E}\{s(k)|t_A(k)\},
\end{align}
where
\begin{align}\label{eq:tAk}
  &t_A(k)\doteq y_{A,B}(k)-h_{B,A}x_A(k) \notag\\
  &= s(k)+w_B(k)+v_A(k).
\end{align}
Here $t_r(k)$ is the sufficient statistics at Alice for $s(k)$. The expression of \eqref{eq:tAk} is an effective return channel from Bob to Alice, which is an additive white Gaussian noise (AWGN) channel.  The SNR of this channel (for $\sigma_B^2\gg \epsilon_A$) is
\begin{equation}\label{}
  \texttt{SNR}_{A|B}= \frac{\sigma_s^2}{\sigma_B^2}.
\end{equation}

\subsection{Effective return channel from Bob to Eve}

The MMSE estimate of $s(k)$ from $\{\mathbf{e}_A(k),y_{E,B}(k),h_{B,A},\mathbf{g}_A,\forall k\}$ is
\begin{align}\label{}
  &\hat s_E(k) = \mathbb{E}\{s(k)|\mathbf{e}_A(k),y_{E,B}(k),h_{B,A},\mathbf{g}_A,\forall k\}
  \notag\\
  &=\mathbb{E}\{s(k)|t_E(k)\},
\end{align}
where
\begin{align}\label{eq:tEk}
  &t_E(k)\doteq y_{E,B}(k)-h_{B,A}\hat x_A(k) \notag\\
  &= s(k)+h_{B,A}\Delta x_A(k)+w_B(k)+v_E(k).
\end{align}
Here $\hat x_A(k)$ is the MMSE estimate of $x_A(k)$ from $\{\mathbf{e}_A(k),\mathbf{g}_A, \forall k\}$, and
$\Delta x_A(k)=x_A(k)-\hat x_A(k)$ is the MMSE estimation error of $x_A(k)$. Specifically, the MSE of $\Delta x_A(k)$
is given by $r_{\Delta x_A}$ in \eqref{eq:rDxAE}.

We see that \eqref{eq:tEk} represents an effective (AWGN) return channel from Bob to Eve where the channel SNR (for $\sigma_B^2\gg \epsilon_E$) is
\begin{align}\label{}
  &\texttt{SNR}_{E|B} = \frac{\sigma_s^2}{|h_{B,A}|^2r_{\Delta x_A}+\sigma_B^2}\notag\\
  &=\frac{\sigma_s^2/\sigma_B^2}{
  \phi_{B,A}+1}<\texttt{SNR}_{A|B},
\end{align}
where $\phi_{B,A}$ is given in \eqref{eq:phiB}.
We also see that $\texttt{SNR}_{E|B}$ is a decreasing function of $p_A$. If  $p_A\|\mathbf{g}_A\|^2/\sigma_{E,A}^2\gg 1$ and $|h_{B,A}|^2/\sigma_B^2=\|\mathbf{g}_A\|^2/\sigma_{E,A}^2$, then $\texttt{SNR}_{E|B}=\frac{1}{2}\texttt{SNR}_{A|B}$.

Also note that $w_B(k)$ is a common noise component in both the legitimate return channel and Eve's return channel. But this feature does not change the achievable secrecy rate $\tilde\xi_{B,A}'$ (in bits per sample) of the WTC from Bob to Alice and Eve. This is because
\begin{equation}\label{}
  \tilde\xi_{B,A}'= I(s(k);t_A(k))-I(s(k);t_E(k)).
\end{equation}
Here we see that whether a noise component in $t_A(k)$ is shared in $t_E(k)$ does not matter since $I(s(k);t_A(k))$ and $I(s(k);t_E(k))$ are computed separately. It follows that
\begin{align}\label{eq:txiBAP}
&  \tilde\xi_{B,A}'=\log(1+\texttt{SNR}_{A|B})-\log(1+\texttt{SNR}_{E|B})\notag\\
&=\log\frac{1+\sigma_s^2/\sigma_B^2}{1+\frac{\sigma_s^2/\sigma_B^2}{\phi_{B,A}+1}}
\notag\\
&=\log\left (1+\frac{\phi_{B,A}\sigma_s^2/\sigma_B^2}{\sigma_s^2/\sigma_B^2+1+\phi_{B,A}}\right )\notag\\
&\leq  \log (1+\phi_{B,A}),
\end{align}
where the upper bound is approached if  $\sigma_s^2/\sigma_B^2\gg 1+\phi_{B,A}$. Also note that the expectation of $\tilde\xi_{B,A}'$ over the distributions of the probing channels equals to $\xi_{B,A}'$ shown in \eqref{eq:xiBAP}.

Hence for a large $\sigma_s^2$,
\begin{equation}\label{eq:EtxiBAP}
  \mathbb{E}\{\tilde\xi_{B,A}'\}=\mathbb{E}\{\log (1+\phi_{B,A})\}
\end{equation}
which is also the upper bound of the secret-key capacity from the original data sets immediately following the channel probing.

It is clear that $\phi_{B,A}$ in \eqref{eq:EtxiBAP} and \eqref{eq:phiB} is $\frac{\texttt{SNR}_{B,A}}{1+\texttt{SNR}_{E,A}}$ shown in \eqref{eq:xi} where $\texttt{SNR}_{B,A}=p_A|h_{B,A}|^2/\sigma_B^2$ and $\texttt{SNR}_{E,A}=p_A\|\mathbf{g}_A\|^2/\sigma_{E,A}^2$.

\subsection{A remark on implementation}

We now see that \eqref{eq:tAk} and \eqref{eq:tEk} represent an \emph{effective} AWGN WTC model from Bob to Alice where the \emph{effective} eavesdropping channel from Bob to Eve is always weaker than the \emph{effective} main channel from Bob to Alice. A best way to utilize this model for a finite $m_A$ is perhaps what is shown in \cite{Yang2019}, which transmits a secret (at a rate close to $\tilde\xi_{B,A}'$) directly from Bob to Alice over the effective WTC model.

Alternatively,  one can use a channel coding method that allows a reliable transmission of $S$ from Bob to Alice at a rate (in bits per sample of $s(k)$) close to $\log(1+\texttt{SNR}_{A|B})$. After this transmission is completed, Alice and Bob will apply a hash function to compress $S$ to a key of size no larger than $m_A\tilde\xi_{B,A,\texttt{min}}'$ bits per realization of $X_A$. Here $\tilde\xi_{B,A,\texttt{min}}'$ is a known (positive) lower bound of $\tilde\xi_{B,A}'$. The last step of using a hash function is also known as privacy amplification.

The classic channel coding methods are available in \cite{Lin2004}. A latest development of channel coding is shown in \cite{Duffy2023}. For hash functions, see \cite{Stinson1994}.

Both of the above approaches are asymptotically optimal (i.e., as the packet length increases). But for finite-length packets, the second approach may have a drawback. Namely, if the transmission rate over the effective return channel is not close enough to $\log(1+\texttt{SNR}_{A|B})$ and falls below $\log(1+\texttt{SNR}_{E|B})$, the secrecy could be compromised (as Eve could have a more powerful decoder than Alice).

\subsection{Power Consumption}
Using STEEP over analog channels, the return signal sent by Bob is
$r(k)=y_B(k)+s(k)$ which consumes the power equal to (strictly speaking proportional to) $p_r\doteq |h_{B,A}|^2p_A+\sigma_B^2+\sigma_s^2=\sigma_B^2(\sigma_s^2/\sigma_B^2+|h_{B,A}|^2p_A/\sigma_B^2+1)$. In order for  $\tilde\xi_{B,A}'$ to be close to its upper bound (see \eqref{eq:txiBAP}), we need $\sigma_s^2/\sigma_B^2\gg 1+\phi_{B,A}$, i.e., $\sigma_s^2/\sigma_B^2\gg 1+\frac{p_A|h_{B,A}|^2/\sigma_B^2}{p_A\|\mathbf{g}_A\|^2/\sigma_{E,A}^2+1}$,
which becomes invariant to $p_A$ if $p_A\|\mathbf{g}_A\|^2/\sigma_{E,A}^2\gg 1$ (i.e., Eve's probing channel has a high SNR). Subject to the above (mild) condition, we have $p_r=|h_{B,A}|^2p_A+\sigma_s^2$ where both terms (after a required power amplification) must be much larger than the power of the noise in the return channel from Bob to Alice. Also, both $h_{B,A}x_A(k)$ and $s(k)$ inside $r(k)$ are important signal components to be received by Alice. Hence, a reasonable choice of $\sigma_s^2$ is such that $|h_{B,A}|^2p_A=\sigma_s^2$ and then $p_r=2\sigma_s^2$ (i.e., 3dB beyond $\sigma_s^2$).

The power consumption is primarily an issue at the physical layer. If the probing channels and return channels are at the link layer or above, then all symbols are digital. As shown next, the extra 3dB power consumption is no longer necessary.

\section{STEEP for Digital Channels}\label{sec:STEEP_digital}

The previous discussions are all focused on analog probing and return channels. However the same principle of STEEP applies to digital probing and return channels as well.

\begin{figure}
  \centering
  \includegraphics[width=2.8in]{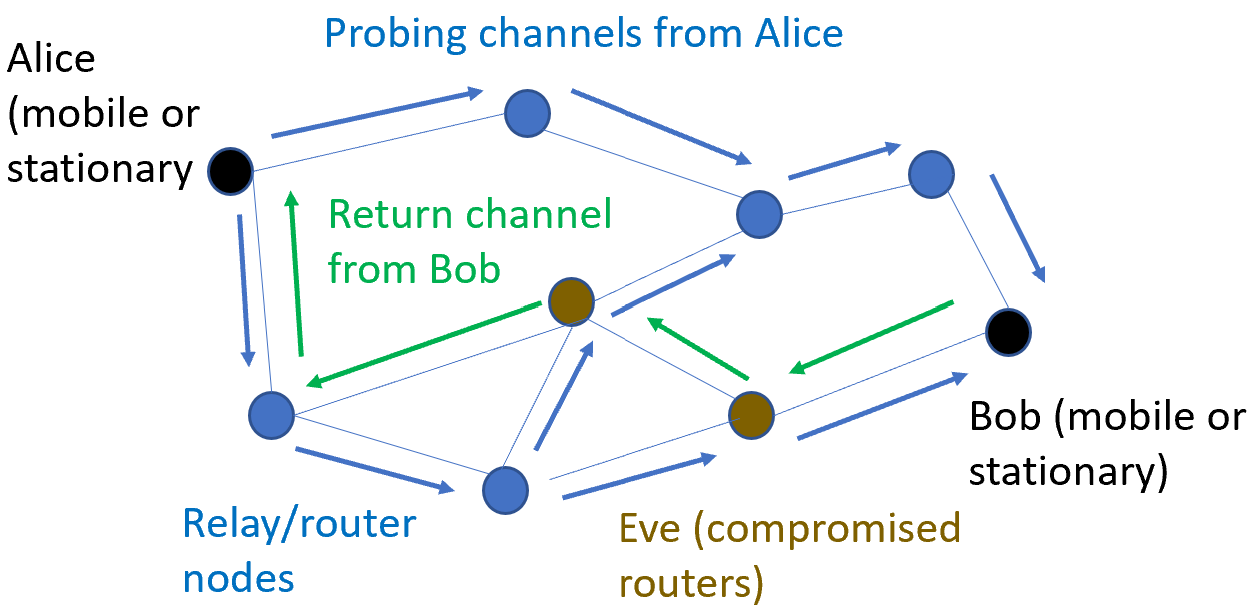}
  \caption{A potential application of STEEP in multihop digital networks.}\label{fig:multihop}
\end{figure}

Now we assume that every channel is digital. A digital channel between two nodes in a network could  be of one or multiple hops.  We can also generalize a channel probing session to include multiple locations of mobile Alice and/or Bob and/or Eve in the network so that different networking routes may be explicitly or implicitly used during the probing session. See Fig. \ref{fig:multihop}. At the network layer, a packet is often either received or lost. If it is received, then all bits in the packet are known to the receiver and transmitter. If it is lost, then each bit in the transmitted packet  has an error rate at $1/2$. After a channel probing session is completed, Alice and Bob can apply a common function to randomize the ordering of all received as well as lost bits. Then each bit in the entire sequence of bits has an effective bit error rate less (usually much less) than $1/2$. We assume that Eve and Bob apply the same process to reorder their bits.

Note that in this paper we assume that Alice and Bob can authenticate the data received from each other, and Eve is unable to destroy the data authentication process. In applications, authentication itself may require a secret key between two parties. In this case, the use of STEEP can help to refresh and/or generate new secret keys while a pre-existing secret key is used for current authentication purpose (before it expires or becomes exposed).

We can now assume that for every bit $b_A(k)$ transmitted by Alice (which is an i.i.d. binary symmetric sequence only known to Alice), Bob and Eve receive respectively
\begin{equation}\label{}
  b_{B,A}(k)=b_A(k)\oplus w_{B,A}(k),
\end{equation}
\begin{equation}\label{}
  b_{E,A}(k)=b_A(k)\oplus w_{E,A}(k).
\end{equation}
  Here all quantities are binary, $\oplus$ is Exclusive-OR, and $k=1,\cdots,m_A$ is the current index/position of a bit.  The bit error rate of the probing channel from Alice to Bob is $P_{B,A}=\texttt{Prob}(b_{B,A}(k)\neq b_A(k))=\texttt{Prob}(w_{B,A}(k)=1)$, and that from Alice to Eve is $P_{E,A}=\texttt{Prob}(b_{E,A}(k)\neq b_A(k))=\texttt{Prob}(w_{E,A}=1)$. For STEEP to have a positive secrecy that increases with $m_A$, we will need $P_{E,A}>0$.

  We will drop the index $k$ for simpler notations.

Then, over a return channel to Alice (and another return channel to Eve), Bob sends $b_r=b_s\oplus b_{B,A}$ (for all $k$) where $b_s$ is an i.i.d. binary symmetric sequence only known to Bob.  Hence, Alice and Eve receive
\begin{align}\label{}
 & b_{A,B}=b_r\oplus w_{A,B} = b_s\oplus b_{B,A}\oplus w_{A,B}\notag\\
 &=b_s\oplus b_A\oplus w_{B,A}\oplus w_{A,B},
\end{align}
\begin{align}\label{}
  &b_{E,B}=b_r\oplus w_{E,B}=b_s\oplus b_{B,A}\oplus w_{E,B}\notag\\
  &=
  b_s\oplus b_A\oplus w_{B,A}\oplus w_{E,B},
\end{align}
where the return channel error rate for Alice is $P_{A,B}=\texttt{Prob}\{w_{A,B}=1\}$, and that for Eve is $P_{E,B}=\texttt{Prob}\{w_{E,B}=1\}$.

Note that Alice can compute
\begin{equation}\label{eq:bbAB}
  \bar b_{A,B}\doteq b_{A,B}\oplus b_A=b_s\oplus w_{B,A}\oplus w_{A,B},
\end{equation}
and Eve can compute
\begin{equation}\label{eq:bbEB}
  \bar b_{E,B}\doteq b_{E,B}\oplus b_{E,A} = b_s\oplus w_{E,A}\oplus w_{B,A}\oplus w_{E,B}.
\end{equation}

Assume that the return channels from Bob to Alice and Eve are much less noisy than the probing channel from Alice to Bob, i.e., $P_{A,B}\ll P_{B,A}<1/2$ and $P_{E,B}\ll P_{B,A}<1/2$. Then
it follows that the effective error rate at Alice about $b_s$ is
\begin{equation}\label{}
  P_{A|B}\doteq \texttt{Prob}(\bar b_{A,B}\neq b_s) = P_{B,A},
\end{equation}
and that at Eve about $b_s$ is
\begin{align}\label{}
  &P_{E|B}\doteq \texttt{Prob}(\bar b_{E,B}\neq b_s)\notag\\
  &=1-P_{B,A}P_{E,A}-(1-P_{B,A})(1-P_{E,A})\notag\\
  &=P_{B,A}+P_{E,A}(1-2P_{B,A})>P_{A|B}.
\end{align}

We see that the effective return channel for Alice (represented by \eqref{eq:bbAB}) is stronger than that for Eve (represented by \eqref{eq:bbEB}). The secrecy capacity (in bits per return sample) of the effective WTC model  is
\begin{align}\label{}
  &\xi \doteq I(b_s;\bar b_{A,B})-I(b_s;\bar b_{E,B})\notag\\
  &=H(\bar b_{E,B}|b_s)-H(\bar b_{A,B}|b_s)\notag\\
  &=f(P_{E|B})-f(P_{A|B}),
\end{align}
with $f(p)=-p\log p -(1-p)\log(1-p)$. Here we have used $H(\bar b_{A,B})=H(\bar b_{E,B})=1$. The (binary) uniform distribution of $b_s$ makes each of $\bar b_{A,B}$ and $\bar b_{E,B}$ (binary) uniformly distributed. It is clear that subject to $P_{E|B}<1/2$, we have $\xi>0$ because of $P_{E|B}>P_{A|B}$.

We show next that there is no loss of secrecy in the above WTC based treatment of the return channels.

After the channel probing from Alice to Bob, the data sets at Alice, Bob and Eve are $\mathcal{A}=\{b_A(k),\forall k\}$, $\mathcal{B}=\{b_B(k),\forall k\}$, and $\mathcal{E}=\{b_{E,A}(k),\forall k\}$. Then according to MAC's bounds in \eqref{eq:MAC}, the secret-key capacity in bits per sample is lower bounded by
\begin{align}\label{}
  &\xi_L\doteq \frac{1}{m_A}C_B= I(b_A;b_B)-I(b_B;b_{E,A})\notag\\
  &=H(b_{E,A}|b_B)-H(b_B|b_A)\notag\\
  &=f(P_{E|B})-f(P_{A|B}),
\end{align}
where we have used $H(b_{E,A})=H(b_B)=1$.  The justification of $H(b_{E,A}|b_B)=f(P_{E|B})$ is that
$b_{E,A}=b_A\oplus w_{E,A}=b_B\oplus w_{B,A}\oplus w_{E,A}$. Here $\texttt{Prob}(b_{E,A}\neq b_B)=\texttt{Prob}(w_{B,A}\oplus w_{E,A}=1)=P_{E|B}$.

Also according to MAC's bounds, the secret-key capacity in bits per sample is upper bound by
\begin{align}\label{}
  &\xi_U\doteq\frac{1}{m_A}C_E= I(b_A;b_B|b_{E,A})\notag\\
  &=H(b_B|b_{E,A})-H(b_B|b_A,b_{E,A}).
\end{align}
Here $H(b_B|b_{E,A})=H(b_{E,A}|b_B)+H(b_B)-H(b_{E,A})=H(b_{E,A}|b_B)=f(P_{E|B})$. For $H(b_B|b_A,b_{E,A})$, recall $b_B=b_A\oplus w_{B,A}$ and $b_{E,A}=b_A\oplus w_{E,A}$, i.e., $b_{E,A}$ is independent of $b_B$ when conditioned by $b_A$. Hence, $H(b_B|b_A,b_{E,A})=H(b_B|b_A)=f(P_{A|B})$.
So, we have
\begin{equation}\label{}
  \xi_U=f(P_{E|B})-f(P_{A|B}).
\end{equation}

Hence,
$\xi_U=\xi_L=\xi$.
Namely, after the channel probing from Alice to Bob, the WTC treatment of the equivalent return channel from Bob to Alice is optimal.

Unlike the case of using analog channels, using digital channels does not cause additional power consumption to establish an effective WTC model such that the effective main channel is guaranteed to be stronger than the effective eavesdropping channel.

Note that for  digital channels, \eqref{eq:xi} should be replaced by $\xi_{\texttt{STEEP},\texttt{DC}}\doteq\xi=f(P_{E|B})-f(P_{A|B})$.

\subsection{A remark on implementation}
The effective WTC model is discrete and memoryless, for which the method in \cite{Wyner1975} is directly applicable to securely transmit a secret (at a rate close to $\xi$)  from Bob to Alice.

However, one can also use a channel coding method to reliably transmit $\{b_s(k),\forall k\}$ (at a rate close to $I(b_s;\bar b_{A,B})=1-f(P_{A|B})$)  over the effective return channel from Bob to Alice. After this transmission is completed, both Alice and Bob apply a hash function to compress the information in $\{b_s(k),\forall k\}$ to a key of a size no larger than $\xi$.

Similar to an earlier comment, the second approach can not guarantee the secrecy if the actual transmission rate from Bob to Alice falls below $I(b_s;\bar b_{E,B})=1-f(P_{E|B})$.

\section{Conclusion}

As a further development from the prior work \cite{Hua2023},  this paper has focused on SISO probing channels between users, and presented important new insights into MAC's bounds. These insights have led to the formulation of STEEP, which has attractive properties for real-world applications. Provided that Eve is unable to receive the exact probes sent by Alice during channel probing, STEEP guarantees a positive secrecy rate (in bits per sample) via any high-quality return channel. This is the case even if the probing channel state information of the users is public. STEEP is applicable to all layers of networks, and it is built on a foundation supported by MAC's bounds for SKG shown in \cite{Maurer1993} and \cite{Ahlswede1993}, and the established methods and theory for WTC such as in \cite{Wyner1975}, \cite{Yang2019} and \cite{Bloch2011}.

\end{document}